\documentclass[11pt,onecolumn,floatfix,altaffilletter,superscriptaddress,tightenlines,showpacs,showkeys,preprintnumbers,nofootinbib]{revtex4-1}
\pdfoutput=1
\usepackage[colorlinks=true,citecolor=blue,linkcolor=blue,breaklinks=true]{hyperref}
\usepackage{amsmath,amssymb}
\usepackage{epsfig} 
\usepackage{graphicx}        
\usepackage{url}
\usepackage{color}
\usepackage{multirow}
\usepackage{placeins}
\usepackage[dvipsnames]{xcolor}
\usepackage{braket}
\usepackage{float}
\usepackage{slashed}
\hypersetup{
  colorlinks=true,
  linkcolor=red,
  filecolor=magenta,   
  urlcolor=blue,
  citecolor=blue,
}

\allowdisplaybreaks

\bibpunct{[}{]}{,}{n}{}{,}

\def\beq{\begin{equation}}
\def\eeq{\end{equation}}
\def\bea{\begin{eqnarray}}
\def\eea{\end{eqnarray}}
\makeatletter
\@addtoreset{equation}{section}

\begin{document}

%\title{Probing Type-I seesaw with curvature-dependent lepton chemical potential}
%\title{Flavour effects in gravitational leptogenesis}
%\title{Importance of flavour effects in right  handed neutrino induced gravitational leptogenesis}
\title{ Gravitational Waves-Tomography of Low-Scale-Leptogenesis}
\author{Satyabrata Datta}
\email{satyabrata.datta@saha.ac.in}
\affiliation{Saha Institute of Nuclear Physics, 1/AF, Bidhannagar, Kolkata 700064, India.}
\affiliation{Homi Bhabha National Institute, 2nd floor, BARC Training School Complex,
	Anushaktinagar, Mumbai, Maharashtra 400094, India.}
%\author{Avik Paul}
%\email{avik.paul@saha.ac.in}
%\affiliation{Astroparticle Physics and Cosmology Division, Saha Institute of Nuclear Physics, HBNI 1/AF Bidhannagar, Kolkata 700064, India}
\author{Rome Samanta}
%\author{Constantinos Skordis}
%\author{Federico R. Urban}
\email{romesamanta@gmail.com}
\affiliation{CEICO, Institute of Physics of the Czech Academy of Sciences, Na Slovance 1999/2, 182 21 Prague 8, Czech Republic}

%\preprint{NUHEP-TH/19-08}

\begin{abstract} 
A  long-lived scalar field ($\Phi$) which couples weakly to the right-handed (RH) neutrinos ($N_{Ri}$), generates small RH neutrino masses ($M_i$) in Low-Scale-Leptogenesis (LSL) mechanisms, despite having a large vacuum expectation value $v_\Phi$. In this case, the correlation shared by the $M_i$s and the duration of the non-standard cosmic history driven by the $\Phi$ provides an excellent opportunity to study LSL signatures on primordial gravitational waves (GWs). We find it engaging, specifically for the gravitational waves that originate due to the inflationary blue-tilted tensor power spectrum and propagate through the non-standard cosmic epoch. Depending on $M_i$, broadly, the scenario has two significant consequences. First, if LSL is at play, GWs with a sizeable blue tilt do not contradict the Big-Bang-Nucleosynthesis (BBN) bound even for the post-inflationary models with very high-scale reheating. Second, it opens up a possibility to probe LSLs via a low-frequency and a complementary high-frequency measurement of  GW-spectral shapes which are typically double-peaked. For a case study, we consider the recent results on GWs from the Pulsar-Timing-Arrays (PTAs) as a `measurement' at the low frequencies and forecast the signatures of  LSL mechanisms at the higher frequencies. 
\end{abstract}

\maketitle
\section{introduction}
Baryogenesis via leptogenesis \cite{lep1} is a two-step process. First, a lepton asymmetry is created, which at the final step gets converted to baryon asymmetry by Sphalerons \cite{Kuzmin:1985mm}. The most straightforward extension of the Standard Model (SM) with three right-handed (RH) neutrinos facilitates light active neutrino masses and leptogenesis. The state-of-the-art RH neutrino mass window corresponding to a successful leptogenesis spans a wide range: from $10^{15}$ GeV down to a few MeV \cite{lep2,lep3,lep4,lep5,lep6,lep7,lep8,lep9,lep10,lep11,lep12}. Nonetheless, the Electroweak naturalness condition  puts an upper bound on the RH neutrino masses: $M_i\lesssim 10^7$ GeV \cite{nat1,nat2}, favoring low-scale-leptogenesis (LSL). Furthermore, because terrestrial experiments such as LHC can reach the energy scale only up to a few TeV,  leptogenesis mechanisms with smaller RH neutrino masses $M_i\lesssim $ TeV have better experimental prospects. \\

In this article, we  find Gravitational Waves (GW) signatures of LSL mechanisms that may complement the low-energy particle physics experiments. After the discovery of GWs from black hole mergers by the LIGO and Virgo collaboration \cite{gw1,gw2}, we are in a new cosmic frontier where experiments are in operation, or plenty of them are being planned to detect GWs from the early universe. Perhaps, we already have a signal because the Pulsar Timing Arrays (PTAs) recently reported strong evidence of a stochastic GW-alike process
 \cite{nanograv,ppta,epta}. \\

The most natural source of primordial GWs is cosmic inflation \cite{inf1,inf2}. Amplitudes of the GWs from the simplest single-field slow-roll inflation are nearly scale-invariant and not large enough to be detectable within the sensitivity range of the present and the planned detectors \cite{inf3}. Nonetheless, many models go beyond the simplest one and predict enhanced or blue-tilted GWs (BGWs) \cite{bgw1,bgw2,bgw3,bgw4,bgw5,bgw6,bgw6,bgw7,bgw8} on small scales, which are detectable. In fact, the possibility that the findings of the PTAs are due to BGWs has been studied in Refs.\cite{bn1,bn2,bn3,bn4}. Moreover, recent works within the swampland conjecture motivate going beyond the single-field slow-roll scenarios in order to achieve a quantum gravity-consistent UV completion \cite{smp1,smp2,smp3,smp4}. Therefore, both theoretical considerations and experimental prospects make it worthwhile to explore the physics associated with the BGWs. Motivated by all these aspects, we discuss the imprints of LSL mechanisms on the BGWs and show how such GWs, if they exist, can test and constrain the LSL mechanisms with a wide range of RH neutrino masses.\\

Because the origin of lepton asymmetry and the BGWs are independent, the simplest way to relate both is to capitalize on the post-inflationary evolution of the early universe. A useful framework to this end is to consider a scalar field that generates RH neutrino masses and dominates the Universe's energy density \cite{Blasi:2020wpy} affecting the GW propagation. For a fixed vacuum expectation value of the scalar field ($v_\Phi$), which may be large, if the Yukawa couplings determine the required smallness of the  RH neutrino masses as well as the lifetime of the $\Phi$,  then an LSL mechanism with a given $M_i$ gets related to the duration of the $\Phi$-dominated epoch -- this is the central idea of the study\footnote{Let us mention that we are not the first to study BGWs in the context of seesaw. Ref.\cite{Asaka:2020wcr} has discussed it with a long-lived RH neutrino and showed a possible connection of BGWs with the lightest active neutrino mass. In addition, recently, there have been efforts to probe leptogenesis with gravitational waves from various cosmological sources \cite{Blasi:2020wpy,Dror:2019syi,Samanta:2020cdk,Datta:2020bht,Samanta:2021zzk,Samanta:2021mdm,Borah:2022byb,Huang:2022vkf,Barman:2022yos,Dasgupta:2022isg,Borah:2022cdx}.}.\\

Such a dynamic generation of the RH neutrino masses is well motivated from a more fundamental theoretical perspective. The lepton number violation in the mechanism (non-zero mass of the RH neutrinos) can be envisaged as breaking of a gauged $U(1)_{B-L}$ \cite{bml1,bml2,bml3,bml4,bml5} which might be an intermediate residual symmetry in the Grand Unified Theories (GUT) $\rightarrow$ SM breaking. After the $U(1)_{B-L}$ is broken spontaneously,  the $\Phi_{B-L}$ rolls down and oscillates around its true vacuum, dominates the energy density as pressure-less dust (matter), and reheats the Universe again. It has been studied in detail \cite{t1,t2,t3,t4,t5,t6} that such matter/dust domination is associated with entropy production, and it introduces spectral distortion by suppressing the GW amplitudes. We relate such spectral distortions to the RH neutrino mass scale $M_i$ and claim them to be the imprints of LSL mechanisms. \\

With the standard reheating scenario, i.e., inflation followed by the inflaton oscillation leading to matter domination before the first reheating at $T=T_{RH}$, the LSLs are of the following consequences: I) For smaller RH neutrino masses, a large value of spectral index becomes viable without contradicting the bounds from Big-Bang Nucleosynthesis (BBN) and LIGO \cite{Peimbert:2016bdg, LIGOScientific:2016jlg} -- this happens even though the  $T_{RH}$ is very high. II) Typically, a peak-dip-peak signal gets imprinted, and the locations of the peaks depend on the LSL model parameters. As an example of the numerical results, a maximally allowed tensor-to-scalar ratio at CMB scale and a fixed value of the spectral index $n_T\simeq 0.8$ (assuming the PTAs$^{\left[f\sim 10^{-9}\rm Hz\right]}$, e.g., the  NANO-Grav, are correct) would imply that the detectors at the interferometer scales$^{\left[f\sim 25 \rm Hz\right]}$ have the potential to test and constrain LSLs with $M_i\lesssim 10^2$ GeV.\\ 

We show further, how due to the presence of the GUT-motivated $U(1)_{B-L}$ symmetry, GWs from cosmic strings could make the mechanism distinct from any other BGW + intermediate matter-domination scenarios with additional  GW-spectral distortions. In the presence of such GWs, one typically expects a peak-plateau-peak signal instead of a peak-dip-peak one. However, such spectral features show up only for a constrained range of gauge coupling; $g^{\prime}\in \left[ 10^{-4}-5\times 10^{-3}\right]$.\\

The rest of the article is organized as follows: In sec.\ref{s2}, we briefly demonstrate the idea of scalar field domination and find the relevant parameter space. In sec.\ref{s3}, we study the imprints of LSL mechanisms on the blue-tilted GWs. We discuss the cosmic string complementarity in sec.\ref{s4}. Finally, we summarise our results in sec.\ref{s5}.

\section{phase transition and the Scalar field dynamics}\label{s2}

Let us briefly review the phase transition scenario. Phase transition \cite{Linde:1978px,Kibble:1980mv,Quiros:1999jp,Caprini:2015zlo,Hindmarsh:2020hop} happens due to the finite temperature correction to the scalar potential. At a very high temperature, the scalar field sits at $\Phi=0$, and as the temperature drops, the field chooses a potential minimum at $\Phi\neq 0$. At $T\rightarrow 0$, it attains so-called the vacuum expectation value $\Phi=v_{\Phi}$. The way the transition would proceed depends on the structure of the finite temperature potential. 
Let us consider the zero temperature tree-level potential $V(\Phi,0)=-\frac{\mu^2}{2}\Phi^2+\frac{\lambda}{4}\Phi^4$ which leads to the  vacuum expectation value $v_{\Phi}=\frac{\mu}{\sqrt{\lambda}}$.  The temperature-dependent effective potential, which causes the system to restore symmetry at a higher temperature, can be expanded as \cite{Quiros:1999jp,Hindmarsh:2020hop}
\bea
V(\Phi,T)=\frac{\lambda}{4}\Phi^4+D(T^2-T_0^2)\Phi^2-ET\Phi^3,\label{tmpdp}
\eea
 where 
 \bea
 D=\frac{3{ g^\prime}^2+4\lambda}{24},~~E=\frac{3{ g^\prime}^3+{ g^\prime}\lambda+3\lambda^{3/2}}{24 \pi}, ~~ T_0=\frac{ \sqrt{12\lambda}v_{\Phi}}{\sqrt{3{ g^\prime}^2+4\lambda}},
 \eea 
 $g^\prime$ is the gauge coupling\footnote{As mentioned in the introduction, ${\rm SM}\times U(1)_{B-L}$ models are well-motivated examples where the RH neutrinos become massive ($M_i=f_N v_{\Phi_{B-L}}$) dynamically after the spontaneous breaking of $U(1)_{B-L}$ gauge symmetry. Therefore, we  study the scalar field dynamics with the effective potential that involves gauge coupling $g^\prime (g_{B-L})$. For simplicity, we shall not use the subscript $B-L$ any further.}, and we assume that the SM Higgs is sequestered from $\Phi$ at the tree level. The last term in Eq.\ref{tmpdp} gives rise to a potential barrier causing a secondary minimum at $\Phi\neq0$, and at $T=T_c$, the two minima become degenerate. At  $T_0 ~(\lesssim T_c)$, the potential barrier between the two  minima vanishes, and the minimum at $\Phi=0$ becomes a maximum \cite{Quiros:1999jp}.  For the potential given in Eq.\ref{tmpdp}, the critical temperature $T_c$ and the field value $\Phi_c\equiv \Phi \left(T_c\right)$ are calculated as  \cite{Quiros:1999jp,Megevand:2016lpr}
\bea
T_c=T_0\frac{\sqrt{\lambda D}}{\sqrt{\lambda D-E^2}},~~ \Phi_c = \sqrt{\frac{4 D}{\lambda}(T_c^2-T_0^2)}.
\eea
With $E \neq 0$, Eq.\ref{tmpdp} leads to a first-order transition with a strength crudely determined by the order parameter $\Phi_c/T_c$ \cite{Quiros:1999jp}. For $\Phi_c/T_c\ll 1$, the transition is very weakly first-order, which can be  treated as a second-order transition as the potential barrier disappears very quickly. In this case, the  transition dynamics can be described by the rolling of the field $\Phi$ from $\Phi=0$ to $\Phi = v_\Phi$.  We shall work with the values of $\lambda$ and  ${ g^\prime}$ so that  $\Phi_c/T_c\ll 1$ is fulfilled\footnote{Using \rm {CosmoTransitions}  \cite{Wainwright:2011kj}, we numerically computed the nucleation rate $\Gamma_B\sim A(T) e^{-S(T)}$, with $S$ being the Euclidean action, and found that it is always close to zero, i.e.,  the false vacuum volume fraction $f_+={\rm Exp[-I(T)]} \simeq 1$, where $I$ is an integral which attains a significant nonzero value  with large  $\Gamma_B$ \cite{Megevand:2016lpr}. Our finding is consistent with Ref.\cite{Blasi:2020wpy}.}. To this end, we choose $\lambda \simeq{ g^\prime}^3$ and ${ g^\prime} \lesssim 10^{-2}$, which correspond to the order parameter $\Phi_c/T_c\lesssim 0.08$. 
%\bea
%\frac{\Phi_c}{T_c}=\frac{1}{\sqrt{2g}}\sqrt{1-\frac{T_0^2}{T_c^2}}\simeq \frac{1}{2\pi}\frac{1}{\sqrt{4+5 g}}\label{op}
%\eea
%We shall work on small $g(=10^{-4})$ limit,  that translate to 
%\bea
%1-\frac{T_0}{T_c}=6.3\times 10^{-7}, ~~\frac{\Phi_c}{T_c}=0.08 ~~ {\rm for} ~  v_{B-L}=10^{13} ~ {\rm GeV}  \label{ops}
%\eea
The scenario would have been different if the parameter space had been $\lambda\lesssim{ g^\prime}^4$. In that case, the strength of the first-order transition is strong, replicating the usual Electroweak phase transition in the SM, wherein an increased ratio of gauge boson to the Higgs mass ($m_H\ll 125$ GeV) makes the transition stronger \cite{Quiros:1999jp,Hindmarsh:2020hop}. \\

Once the field rolls down to the true vacuum, it oscillates around $v_{\Phi}$. For an arbitrary potential, such oscillation dynamics can be captured with the {\it action-angle formalism}  \cite{aaf1,aaf2,aaf3} in a static universe; the results nonetheless apply to an expanding universe for $\Omega \gg H$, where $\Omega$ is the oscillation frequency and $H$ is the Hubble parameter. 
The total energy density of the scalar field is given by $\rho=\frac{1}{2}\dot{\Phi}^2+V(\Phi)=\mathcal{H}(P,\Phi )$, where $\mathcal{H}$ is the Hamiltonian of the system and $\dot{\Phi}=P$. Here the generalised coordinates are  $P$ and $\Phi$ that  in the {\it action-angle formalism}, map to $I$ and $\theta$ defined as \cite{aaf1} 
\bea
I=\oint P d\Phi =2\int_{\Phi_{\rm min}}^{\Phi_{\rm max}}\sqrt{2\left(\rho-V\right)}d\Phi ~~{\rm and }~~\theta=\frac{dW}{dI}.\label{aav}
\eea
The integration is over an entire period of oscillation, and $W$ is Hamilton's characteristic function that does not explicitly depend on time. The equations of motion are given by \cite{aaf1}   
\bea
\frac{d I}{dt}=\frac{d}{dt}\left(\frac{d\mathcal{L}}{d\dot{\theta}}\right)=-\frac{d\mathcal{K}}{d\theta}=0~~{\rm and}~~\frac{d \theta}{dt}=\frac{d\mathcal{K}}{dI}=\Omega(I), \label{acang}
\eea
where $\mathcal{L}$ and $\mathcal{K}$ are the new Lagrangian and Hamiltonian of the system. In Eq.\ref{acang}, the first relation shows that $I$ is a constant of motion and $\theta$ is a cyclic coordinate (from the Euler-Lagrange equation), implying the new Hamiltonian is only a function of $I$. Therefore, from the second relation, one finds $\Omega(I)$ to be a constant that can be identified as the frequency of motion. This is because, from Eq.\ref{acang}, we have $\theta(t)=\Omega(I)t+c$ with $c$ being the integration constant, and therefore, if the original generalised coordinate undergoes oscillation over a period $T$, the corresponding change in the angle variable becomes $\Delta \theta=\Omega(I) T=1$, since $\Delta\theta=\oint d\Phi\frac{d\theta}{d\Phi} =\frac{d}{dI}\oint \frac{dW}{d\Phi}d\Phi=\frac{d}{dI}\oint Pd\Phi=1$. Hence 
\bea
\Omega(I)=\frac{1}{T}=\frac{d\mathcal{K}}{dI}=\frac{d\rho}{dI}.\label{angv}
\eea
The same method can be applied in an expanding universe provided $\Omega(I)\gg H$.  In that case, we can average over the quantities like $\dot{\Phi}$ to obtain
\bea
\braket{\dot{\Phi}^2}=\frac{1}{T}\oint P\dot{\Phi}dt=I \frac{d\rho}{dI}.
\eea
Consequently, the equation of motion of the scalar field 
\bea
\frac{d\rho}{dt}=-3H\dot{\Phi}^2, 
\eea
leads to $I\propto a^{-3}$, where $H=\dot{a}/a$, with $a$ being the scale factor. The equation of state (E.O.S) parameter $\omega=p/\rho$, where $p=\braket{\dot{\Phi}^2}-\rho$ is the average pressure, is obtained as
\bea
\omega=\frac{I}{\rho}\frac{1}{dI/d\rho}-1. \label{eos}
\eea 
Eq.\ref{eos} is the most important equation in this discussion, because for a constant $\omega$, Eq.\ref{eos} leads to 
\bea \rho \propto I^{(1+\omega)} \propto a^{-3(1+\omega)}.
\eea
Furthermore, given a generic potential  $V(\Phi)=\alpha \Phi^\beta$, from  Eq.\ref{aav} we obtain
\bea
I=\frac{4\sqrt{2\pi}\Gamma\left(\frac{1}{\beta}\right)}{\left(\beta+2\right)\Gamma\left(\frac{1}{\beta}+\frac{1}{2}\right)}\alpha^{-1/\beta}\rho^{\frac{1}{2}+\frac{1}{\beta}}.\label{acv}
\eea
where the integration limits are $V(\Phi_{max})=V(\Phi_{min})=\rho$.
Therefore, Eq.\ref{eos} leads to 
\bea
\omega=(\beta-2)(\beta+2)^{-1}.\label{eosb}
\eea
 We shall assume that the oscillation of the scalar field is driven by the dominant quadratic term in the potential.  The expansion of the zero temperature potential around the true vacuum gives $\alpha=\lambda v_{\Phi}^2$ and $\beta=2$. Therefore, from Eq.\ref{eosb}, one obtains the E.O.S $w=0$ (i.e., the field behaves like matter), and from Eq.\ref{acv} along with Eq.\ref{angv},  one obtains the angular frequency of oscillation $\tilde{\Omega}=2\pi\Omega =\sqrt{2\lambda}v_{\Phi}=m_\Phi$.\\

  The decay channels set the lifetime of the scalar field. Because we  assume $\lambda \simeq {g^\prime} ^3$ and ${g^\prime} \ll 1$, $\Phi\rightarrow Z^\prime Z^\prime $  is not allowed from kinematic consideration. Furthermore, due to the assumed absence of the coupling with the SM Higgs, the main competitive decay channels are $\Phi\rightarrow N_iN_i$ and $\Phi\rightarrow f\bar{f}V$, where $f$ and $V$ are SM fermions and vector bosons. The former is a tree-level process while the latter is a one-loop triangle process. The strength of these two processes is determined by $f_N$ and $g^\prime $ couplings, respectively. Since we want the dynamics to be controlled by $f_N$ (to suppress the RH neutrino masses ($M_i=f_N v_\Phi$) for triggering LSL and at the same time to obtain non-standard cosmological evolution), we shall always work with $f_N$ and $g^\prime$ such that $\Phi\rightarrow N_iN_i$ process dominates ($\Gamma_{N}^\Phi\gtrsim \Gamma_{ffV}$), i.e., this process determines the lifetime of $\Phi$. 
  \\
  
The energy density components in the early universe evolve as \cite{Giudice:2000ex}
  
  \bea
\frac{d\rho_R}{dt}+4H\rho_R=\Gamma_{N}^\Phi\rho_{\Phi},~~
\frac{d\rho_{\Phi}}{dt}+3H\rho_{\Phi}=-\Gamma_{N}^\Phi\rho_{\Phi},~~
\frac{ds}{dt}+3Hs=\Gamma_{N}^\Phi\frac{\rho_{\Phi}}{T},\label{be3}
\eea
where the decay width\footnote{The Yukawa coupling $f_N$ is bounded from above. This is because, one can generate the portal coupling $\lambda_{H \Phi}$ at the radiative level \cite{Brivio:2017dfq}, giving $\lambda_{H \Phi}^{\rm loop}\sim y_D^2 f_N^2$. Considering the light neutrino masses $\sim 0.01$ eV \cite{Esteban:2020cvm} and using $M_i=f_N v_\Phi$, one obtains $f_N\lesssim 3\times 10^{6}/v_\Phi \rm GeV^{-1}$ for $\Phi\rightarrow hh$ not to dominate the $\Phi\rightarrow N_iN_i$ process. On the other hand, the lower bound on $f_N$ is specific to the models of low-scale leptogenesis. For example, the thermal resonant type scenarios would imply $M_i\gtrsim 3\times 10^{2}$ GeV \cite{DiBari:2019zcc} for Sphaleron to fully convert the lepton asymmetry to the $B-L$ asymmetry. In this case, one has $f_N \gtrsim 3\times 10^{2}/v_\Phi \rm GeV^{-1}$. Therefore, for a benchmark value of $v_\Phi=10^{13}$ GeV, the coupling $f_N$ is constrained as $3\times 10^{-11}\lesssim f_N\lesssim 3\times 10^{-7}$.} $\Gamma_{N}^\Phi\sim\frac{f_N^2}{10\pi}m_\Phi$ and Eq.\ref{be3} represent an entropy production equation due to the $\Phi$-decay. It is useful to recast the above equations as
 \bea
\frac{d\rho_{R}}{dz}+\frac{4}{z}\rho_R=0, ~~
\frac{d\rho_{\Phi}}{dz}+\frac{3}{z}\frac{H}{\tilde{H}}\rho_{\Phi}+\Gamma_{N}^\Phi\frac{1}{z\tilde{H}}\rho_{\Phi}=0,\label{den2}
\eea
where $z=M_0/T$ with $M_0$ being an arbitrary mass scale chosen here to be $M_0=T_c$, and from the third of Eq.\ref{be3}, the relation between temperature and time has been derived as
\bea
\frac{1}{T}\frac{dT}{dt}=-\left(H+\frac{1}{3g_{*s}(T)}\frac{dg_{*s}(T)}{dt}-\Gamma_{N}^\Phi\frac{\rho_{\Phi}}{4\rho_{R}}\right)=-\tilde{H}.\label{temvar}
\eea
In the computation, we shall consider $g_*\equiv g_{*s}$ constant. The amount of entropy production can be calculated by solving 
\bea
\frac{da}{dz}=\left(1+\Gamma_N^\Phi\frac{\rho_{\Phi}}{4\rho_{R}\tilde{H}}\right)\frac{a}{z}
\eea
 and computing the ratio ($\mathcal{\kappa}$) of the quantities $S\sim a^3/z^3$ after and before the $\Phi$-dominated epoch. We find an approximate analytical expression for $\mathbb{\kappa}$ as 
 \bea
 \mathbb{\kappa}^{-1}\simeq \frac{\left(\frac{90}{\pi^2 g_*}\right)^{1/4} \rho_{R}\left( T_c \right)\sqrt{\Gamma_N^\Phi \tilde{M}_{Pl}}}{\rho_{\Phi}\left( T_c \right) T_c}\simeq  \frac{3^{1/4}\left(\frac{30}{\pi^2 g_*}\right)^{-3/4} T_c^3\sqrt{\Gamma_N^\Phi \tilde{M}_{Pl}}}{V_{\rm eff}\left( 0,T_c \right)},\label{entrp}
 \eea
 where $\tilde{M}_{Pl}=2.4\times 10^{18}$ GeV is the reduced Planck constant and $\rho_{\Phi}\left( T_c \right) \equiv V_{\text{eff}}\left( 0,T_c\right)\simeq \frac{\lambda}{4}v^4$.\\
 
% To obtain $\rho_{\Phi}\left( T_c \right) \equiv V_{\text{eff}}\left( 0,T_c\right)$, we worked out the full effective potential $V_{\text{eff}}\left( \Phi,T_c\right)=V_{\text{tree}}\left( \Phi\right)+V_{\text{1-loop}}^{\rm CW}\left( \Phi\right)+V_{\text{1-loop}}\left( \Phi,T\right)+V_{\text{ring}}\left( \Phi\right)$ and found that the dominant contribution comes from the temperature dependent part of $V_{\text{1-loop}}\left( \Phi,T\right)$, i.e., $V_{\text{eff}}\left( 0,T_c\right)\equiv V_{\text{1-loop}}\left( 0,T_c\right) \simeq {\pi^2 T_c^4}/{90}$. For the chosen parameter space in this article, $V_{\text{1-loop}}\left( 0,T_c\right)$ always dominates over the vacuum energy of $\Phi$.\\

\begin{table}
	\begin{center}
		\begin{tabular}
			{ |p{2.45cm}||p{2.45cm}|p{2.45cm}|p{2.45cm}|p{2.45cm}|p{2.45cm}|}
			\hline
			\multicolumn{6}{|c|}{Set A: $v_\Phi=10^{13}{\rm GeV}$, $m_\Phi=4.47\times 10^{8}{\rm GeV}$} \\
			\hline
			Parameters & $g^\prime$  & $\Gamma_N^\Phi/{\rm GeV}$ & $T_c/{\rm GeV}$& $M_i/{\rm GeV}$ & $H (T_c)/{\rm GeV}$\\
			\hline
			BP1 & $10^{-3}$  & $1.3\times 10^{-14}$ & $6.3\times 10^{11}$ & 300 & $5.9\times 10^{5}$ \\
			BP2 & $10^{-3}$  & $1.3\times 10^{-16}$ & $6.3\times 10^{11}$ & 30 & $5.9\times 10^{5}$\\
			BP3 & $10^{-3}$  & $1.3\times 10^{-18}$ & $6.3\times 10^{11}$ & 3 & $5.9\times 10^{5}$ \\
			\hline
		\end{tabular}
		\begin{tabular}
			{ |p{2.45cm}||p{2.45cm}|p{2.45cm}|p{2.45cm}|p{2.45cm}|p{2.45cm}|}
			\hline
			\multicolumn{6}{|c|}{Set B: $v_\Phi=10^{13}{\rm GeV}$, $M_i=300{\rm GeV}$} \\
			\hline
			Parameters & $g^\prime$  & $\Gamma_N^\Phi/{\rm GeV}$ & $T_c/{\rm GeV}$& $m_\Phi/{\rm GeV}$ & $H (T_c)/{\rm GeV}$\\
			\hline
			BP1 & $10^{-2}$  & $4.05\times 10^{-13}$ & $2\times 10^{12}$ & $1.4\times 10^{10}$ & $5.9\times 10^{6}$ \\
			BP2 & $10^{-3}$  & $1.3\times 10^{-14}$ & $6.3\times 10^{11}$ & $4.47\times 10^{8}$ & $5.9\times 10^{5}$ \\
			BP3 & $10^{-4}$  & $4.05\times 10^{-16}$ & $2\times 10^{11}$ & $1.4\times 10^{7}$ & $5.9\times 10^{4}$ \\
			\hline
		\end{tabular}
	\end{center}
\caption{Two sets of benchmark points that produce the plots in Fig.\ref{fig1} for $v_\Phi=10^{13}$ GeV. Set A: constant $m_\Phi$ and Set B: constant $M_i$.}\label{t1}
\end{table}

\begin{figure}[H]
\includegraphics[scale=.45]{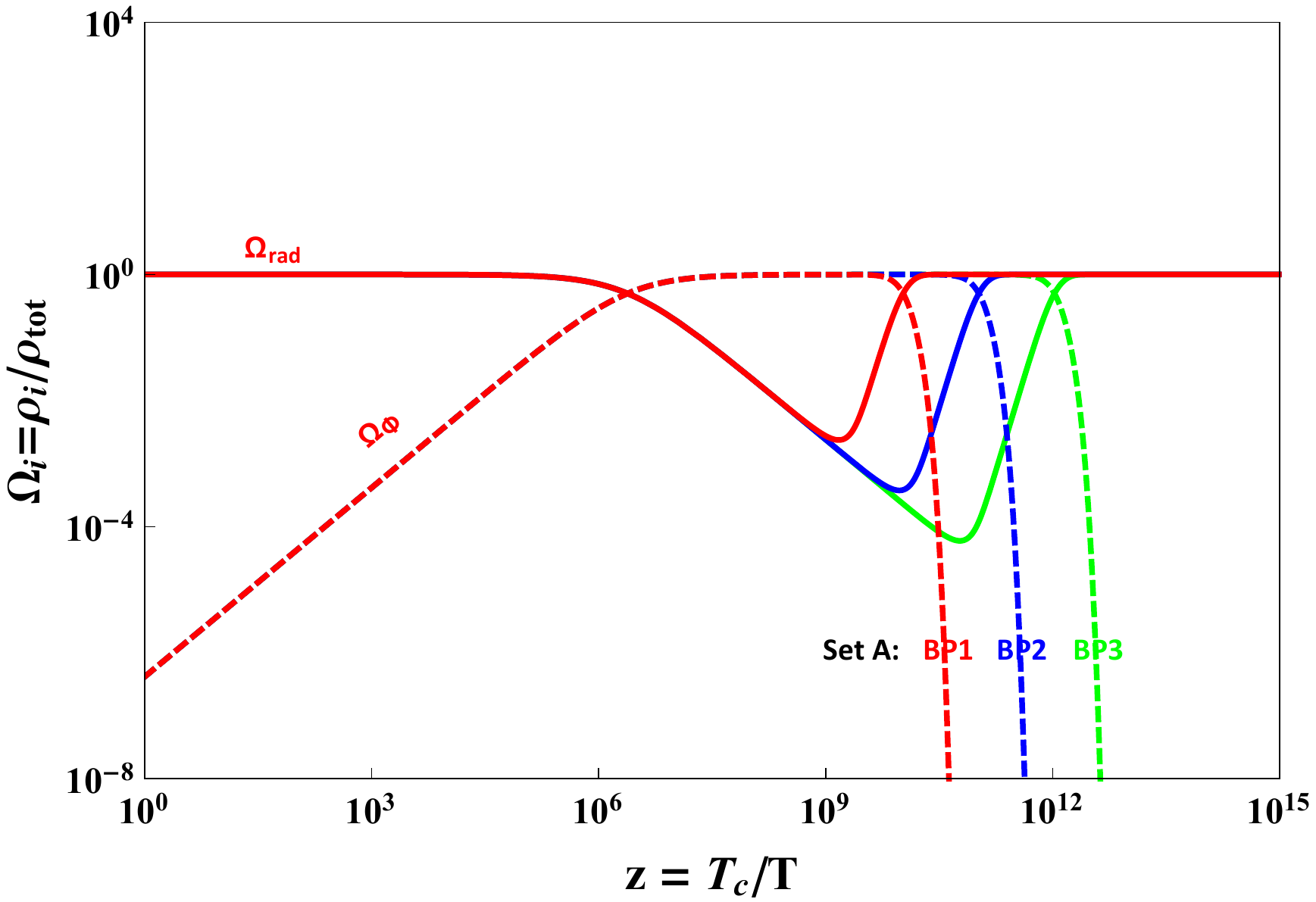} \includegraphics[scale=.45]{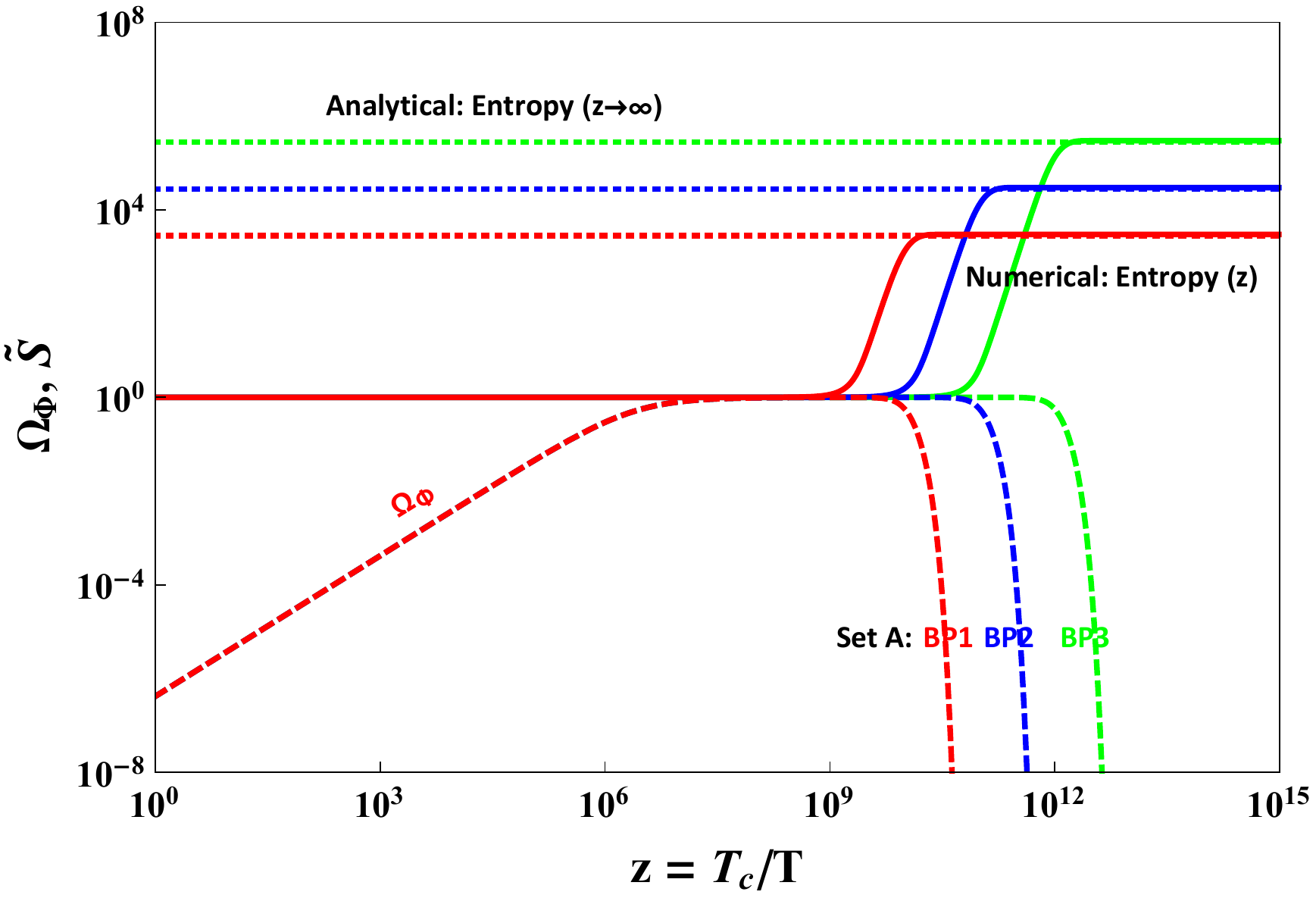} \\
\includegraphics[scale=.45]{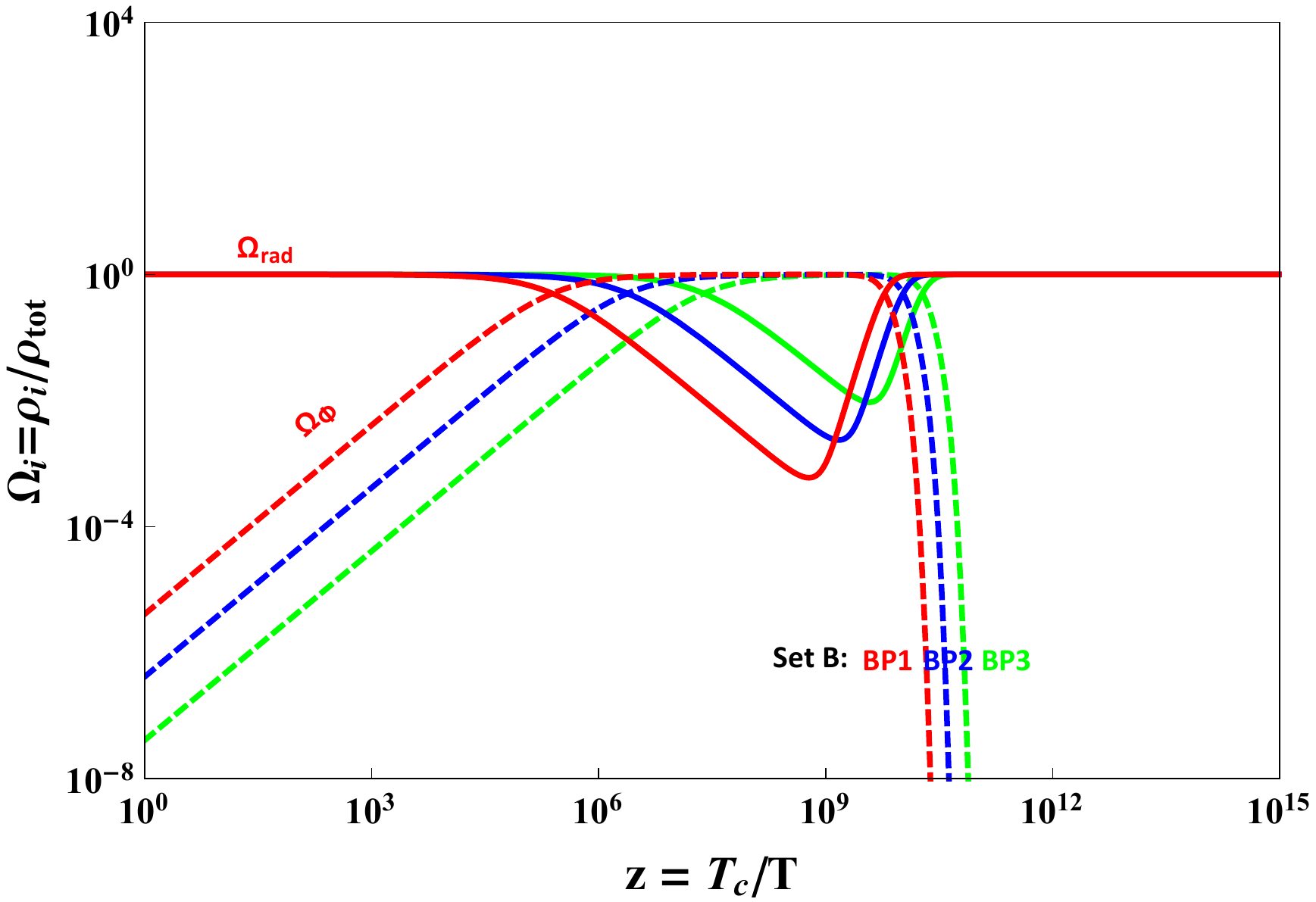} \includegraphics[scale=.45]{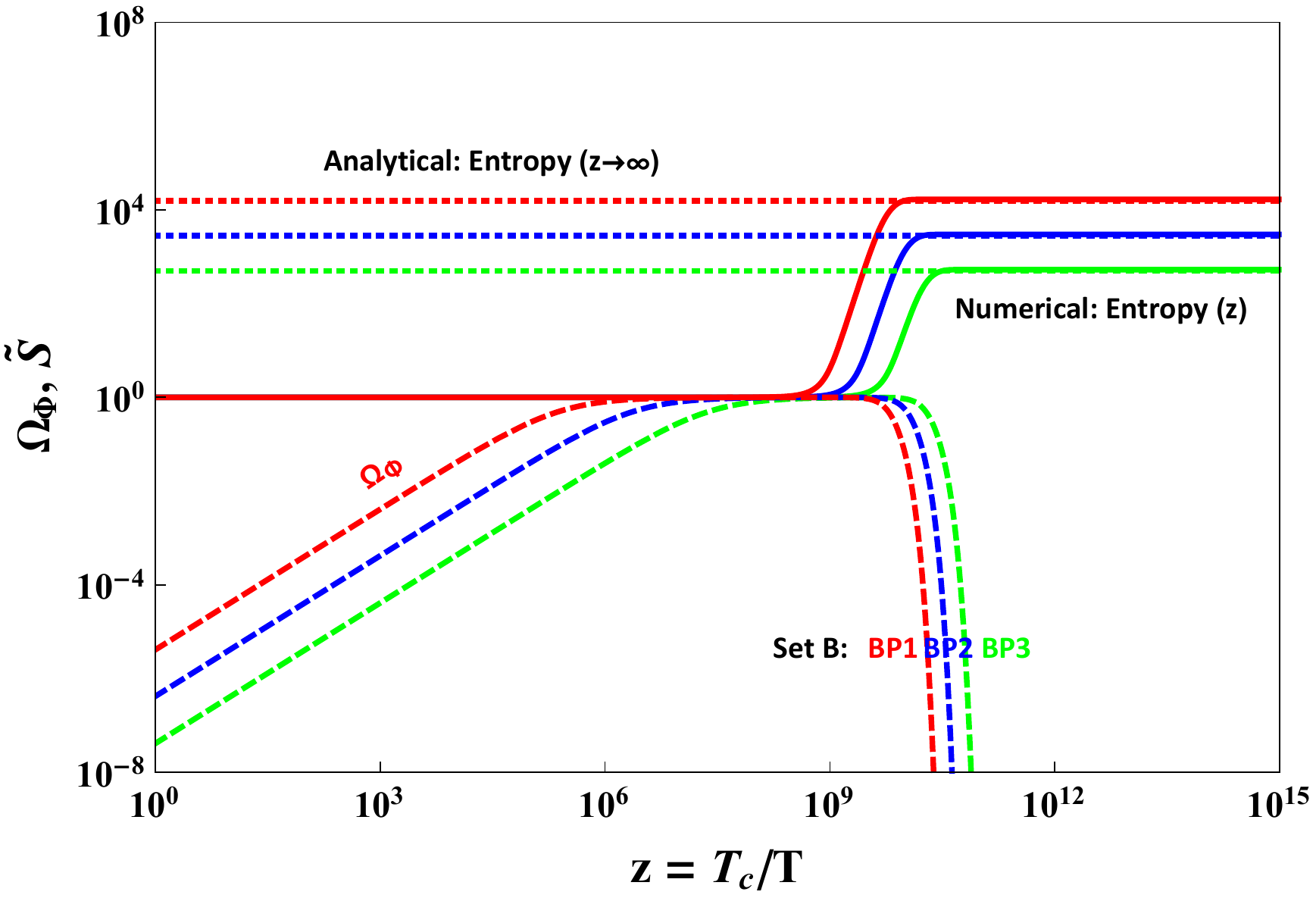} 
\caption{Top panel $\Rightarrow$ Set A and bottom panel $\Rightarrow$ Set B (see Table \ref{t1}):  Left: Evolution of radiation (solid) and the scalar field (dashed) energy densities with the inverse of temperature. Right: Evolution of the scalar field (dashed) energy densities and the total entropy (solid)  with the inverse of temperature. The horizontal lines represent the analytical value of the total entropy at $z\rightarrow\infty$ (see Eq. \ref{entrp}). The total $\tilde{S}$ entropy is normalised such that $\tilde{S}(z=1)$ becomes one. All these figures have been produced for $v_\Phi=10^{13}$ GeV.}\label{fig1}
\end{figure}

In Fig.\ref{fig1},  we show the solutions of the Eq.\ref{den2} for different benchmark values tabulated in Table \ref{t1}. In the upper panel, we show the evolution of the energy densities for $g^\prime=10^{-3}$ and therefore, for a fixed $m_\Phi$ as $\lambda\simeq {g^\prime}^3$.   The solid lines (left) represent the normalized radiation energy densities, whereas the non-horizontal dashed lines do the same for the scalar field on both sides. The solid lines on the right represent the evolution of total entropy $\tilde{S}(z)$ that matches at $z\rightarrow \infty$ (horizontal dashed lines) with the analytical expression obtained in Eq.\ref{entrp}. All the relevant parameters that produce the figures in the upper panel are tabulated in Table \ref{t1} as Set A. The description of the figures in the lower panel would be the same; however, this time, we fix the RH neutrino mass and vary the gauge coupling, consequently the $m_\Phi$. Relevant parameters are listed as Set B in Table \ref{t1}. From Fig.\ref{fig1}, we can extract the following pieces of information. I) The scalar field energy $V_{\rm eff}(0,T_c)$  is always sub-dominant than that of the radiation at $T=T_c$. Therefore, the universe does not go through the second period of inflation. II) Longer the duration of scalar field domination, the larger the entropy production. In the upper panel, a longer duration corresponds to a small RH neutrino mass, whereas, in the lower panel, a stronger gauge coupling does the same.  \\

\begin{figure}
\includegraphics[scale=.6]{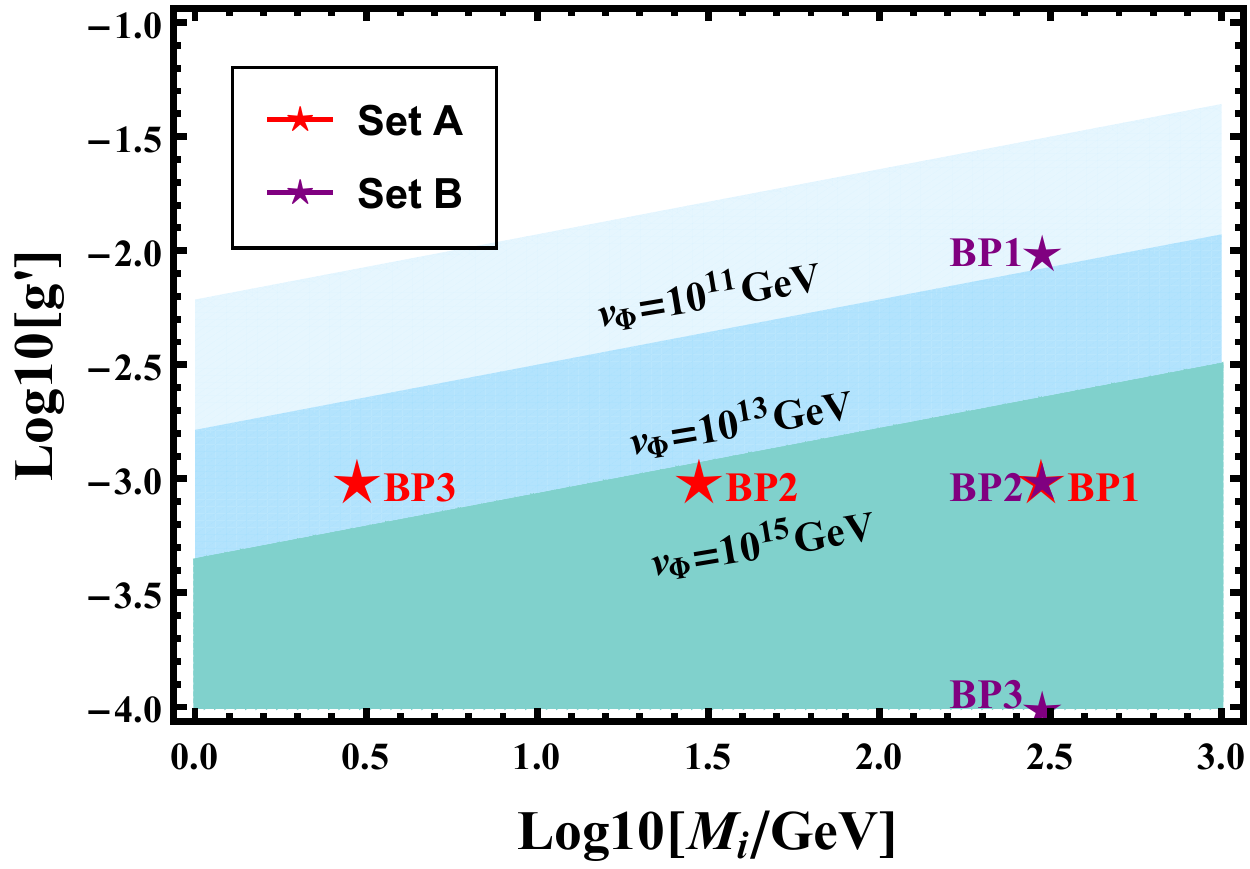}\includegraphics[scale=.8]{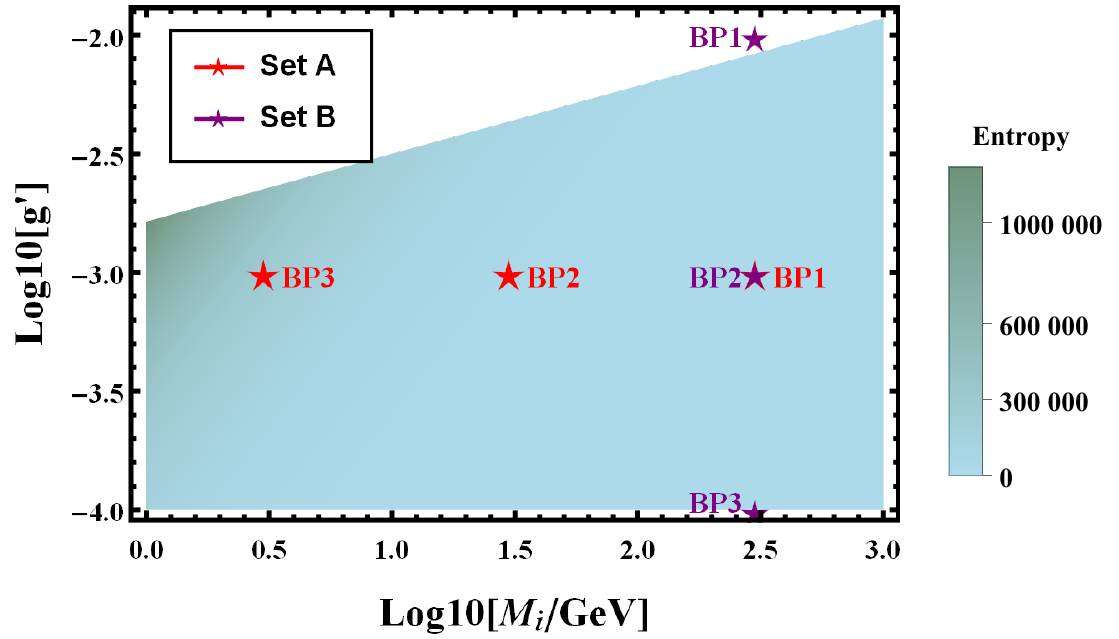}
\caption{Left panel: Allowed parameter space on the $g^\prime-M_i$ plane for the scalar field dynamics to be determined by $\Phi\rightarrow N_iN_i$ decays. The nature of the parameter space for different values of $v_\Phi$ are shown with different colour shades. The red and the purple stars represent the benchmark points in Table \ref{t1}. Right: Density plot of the parameter space on the $g^\prime-M_i$ plane for $v_\Phi=10^{13}$ GeV. The color gradient represents the varying densities of the total entropy.}\label{fig2}
\end{figure}

Let us also point out a subtlety regarding the choice of  gauge coupling for this mechanism to work. We mentioned previously that the competitive decay channel $\Phi\rightarrow ff V$ should be sub-dominant than the $\Phi\rightarrow N N$. In Fig.\ref{fig2} (left), for different values of $v_\Phi$,  we  show  numerically extracted approximate parameter regions  on the $g^\prime-M_i$ plane that comply with $\Gamma_{N}^\Phi\gtrsim \Gamma_{ffV}$. Note that BP1 in Set B is a bad data point for $v_\Phi=10^{13}$ GeV. In the right panel, we show the same parameter space for $v_\Phi=10^{13}$ GeV, but now with a varying color gradient representing the produced entropy for a given set of $g^\prime$ and $M_i$. Though the parameter space shrinks towards small $M_i$ and smaller values of $g^\prime$, in this region, the entropy production is huge -- a fact that can be understood from the right-panel-plots in Fig.\ref{fig1} and has a great significance in the next section's discussions on the GWs.
 \section{ Cosmic archaeology with leptogenesis and tensor blue tilt } \label{s3}
We now briefly review the gravitational waves production during inflation and their  propagation through several cosmic epochs until today. Gravitational waves are described with the perturbed FLRW  line element: 
\bea
ds^2=a(\tau)\left[-d\tau^2+(\delta_{ij}+h_{ij})dx^idx^j)\right],
\eea
where $\tau$ is the conformal time, $a(\tau)$ is the scale factor. The transverse and traceless ($\partial_ih^{ij}=0$, $\delta^{ij}h_{ij}=0$) part of the of the $3\times 3$ symmetric matrix $h_{ij}$ represents the gravitational waves. Because the GWs are weak, $|h_{ij}|\ll1$, the linearized evolution equation 
\bea
\partial_\mu(\sqrt{-g}\partial h_{ij})=16\pi a^2(\tau) \mathcal{\pi}_{ij}\label{lineq}
\eea
 would suffice to study the propagation of the GWs. The quantity $\pi_{ij}$ is the tensor part of the anisotropy stress that couples to $h_{ij}$ as an external source, and in a realistic cosmic setting, it only affects the GW spectrum at scales larger than those of PTAs, e.g., due to neutrino free streaming \cite{Weinberg:2003ur,Zhao:2009we}. It is convenient to express $h_{ij}$ in the Fourier space: 
 \bea
 h_{ij}(\tau, \vec{x})=\sum_\lambda\int \frac{d^3\vec{k}}{(2\pi)^{3/2}} e^{i\vec{k}.\vec{x}}\epsilon_{ij}^\lambda(\vec{k})h_{\vec{k}}^\lambda(\tau),\label{fug}
 \eea
where  the index $\lambda=``+/-"$ represents two polarisation states of the GWs. The polarisation tensors apart from being transverse and traceless,  satisfy the conditions: $\epsilon^{(\lambda)ij}(\vec{k})\epsilon_{ij}^{(\lambda^\prime)}(\vec{k})=2\delta_{\lambda\lambda^\prime}$ and $\epsilon^{(\lambda)}_{ij}(-\vec{k})=\epsilon^{(\lambda)}_{ij}(\vec{k})$. Assuming isotropy and the same evolution of each polarisation state, we rename $h_{\vec{k}}^\lambda(\tau)$ as $h_{k}(\tau)$, where $k=|\vec{k}|=2\pi f$ with $f$ being the frequency of the GWs today at $a_0=1$. With the sub-dominant contribution from $\pi_{ij}$, the GW propagation equation in the  Fourier space becomes
\bea
\ddot{h}_k+2\frac{\dot{a}}{a}\dot{h}_k+k^2h_k=0, \label{prpeq}
\eea
where the dot indicates a conformal time derivative. Using eq.\ref{fug} and Eq.\ref{prpeq}, one calculates the energy density of the GWs  as \cite{WMAP:2006rnx}
\bea
\rho_{GW}=\frac{1}{32\pi G}\int\frac{dk}{k}\left(\frac{k}{a}\right)^2T_T^2(\tau, k)P_T(k),\label{gw1}
\eea
where $T_T^2(\tau, k)=|h_k(\tau)|^2/|h_k(\tau_i)|^2$ is a transfer function which is computed from Eq.\ref{prpeq}, with $\tau_i$ as an initial conformal time. The quantity $P_T(k)=\frac{k^3}{\pi^2}|h_k(\tau_i)|^2$ characterises the primordial power spectrum, which connects to the inflation models with specific forms. Generally, $P_T(k)$ is parametrised as a power-law given by
\bea
P_T(k)=r A_s(k_*)\left(\frac{k}{k_*}\right)^{n_T},
\eea

where $r\lesssim 0.06$ \cite{BICEP2:2018kqh} is the tensor-to-scalar-ratio,  $A_s \simeq 2\times 10^{-9}$ is the scalar perturbation amplitude at the pivot scale $k_*=0.01\rm  Mpc^{-1}$. We treat the tensor spectral index $n_T$ as a constant and blue-tilted ($n_T>0$), although there might be scale dependence owing to the higher order corrections \cite{Kuroyanagi:2011iw}. Let us mention that the single field slow-roll inflation models correspond to the so called consistency relation: $n_T=-r/8$ \cite{Liddle:1993fq}, making the spectral index slightly red-tilted ($n_T\lesssim 0$). The GW energy density  relevant for detection purpose is expressed as 
\bea
\Omega_{GW}(k)=\frac{k}{\rho_c}\frac{d\rho_{GW}}{dk},
\eea
where $\rho_c=3H_0^2/8\pi G$ with   $H_0\simeq 2.2 \times 10^{-4}~\rm Mpc^{-1}$ being the present-day Hubble constant. From Eq.\ref{gw1}, the quantity $\Omega_{GW}(k)$ is derived as 
\bea
\Omega_{GW}(k)=\frac{1}{12H_0^2}\left(\frac{k}{a_0}\right)^2T_T^2(\tau_0,k)P_T(k), ~~\tau_0=1.4\times 10^4 {\rm ~Mpc}.\label{GWeq}
\eea
Several works are dedicated to compute the transfer function analytically \cite{t1,t2,t3,t4}, and we shall use the one reported in Refs.\cite{t5,t6}. In the presence of an intermediate matter domination, $T_T^2(\tau_0,k)$ is given by
\bea
T_T^2(\tau_0,k)=F(k)T_1^2(\zeta_{\rm eq})T_2^2(\zeta_{\Phi})T_3^2(\zeta_{\Phi R})T_2^2(\zeta_{R}),
\eea

where $F(k)$ reads
\bea
F(k)=\Omega_m^2\left( \frac{g_*(T_{k,\rm in})}{g_{*0}}\right)\left(\frac{g_{*s0}}{g_{*s}(T_{k,\rm in})}\right)^{4/3}\left(\frac{3j_1(k\tau_0)}{k\tau_0}\right)^2.\label{fuk}
\eea
\begin{figure}
\includegraphics[scale=.55]{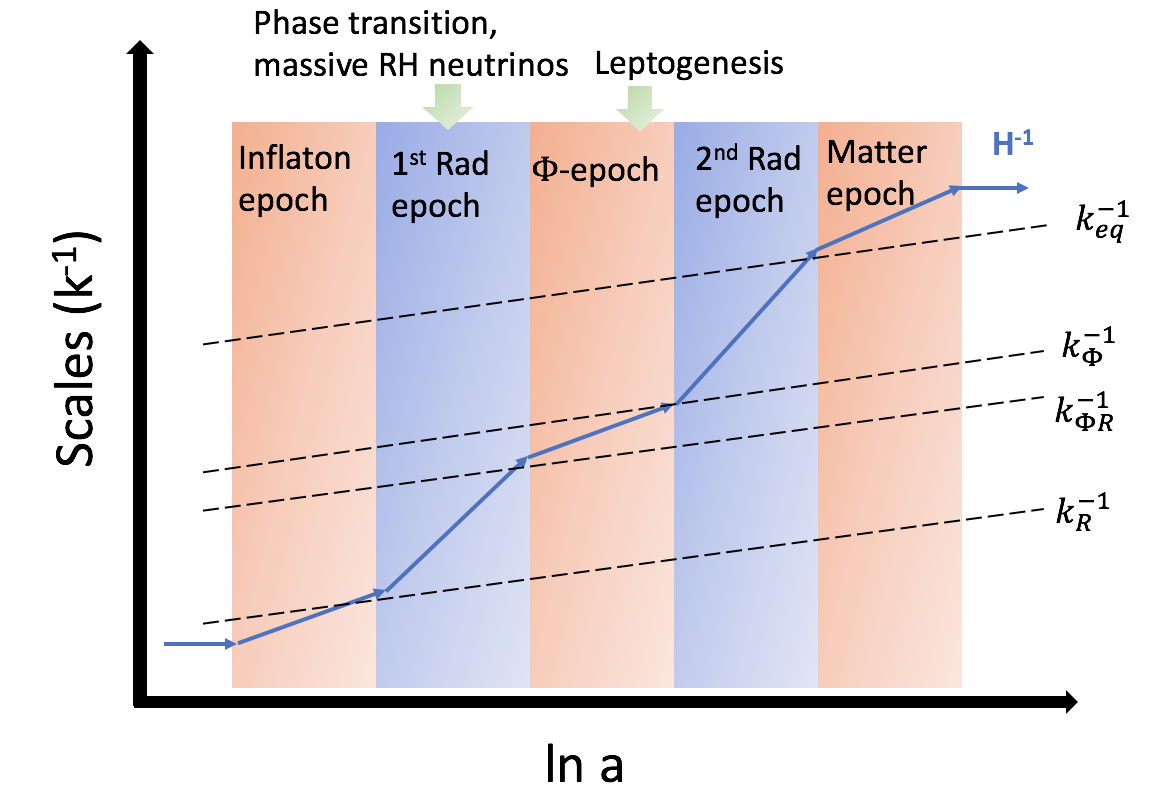}
\caption{A schematic representing the horizon entry of the relevant scales. The orange (blue) color indicates a matter (radiation) domination. {\bf Inflaton epoch:} After the end of inflation, inflaton oscillates and gives matter domination. {\bf 1st Rad epoch:} Inflaton reheats the universe to a radiation domination. {\bf $\Phi$-epoch:} The long-lived scalar field  giving mass to the RH neutrinos leads to a matter domination. {\bf 2nd Rad epoch:} The scalar field reheats the universe to a radiation domination. {\bf Matter epoch:} The standard matter dominated epoch in $\Lambda\rm CDM$ cosmology.}\label{fig3}
\end{figure}
Here $j_1(k\tau_0)$ is the spherical Bessel function, $\Omega_m=0.31$, $g_{*0}=3.36$, $g_{*0s}=3.91$ and an approximate form of the scale-dependent $g_*$ \cite{gs1,gs2,t6} is given in the Appendix \ref{a1}. The transfer functions are given by 
\bea
T_1^2(\zeta)=1+1.57\zeta+ 3.42 \zeta^2,\\
T_2^2(\zeta)=\left(1-0.22\zeta^{1.5}+0.65\zeta^2 \right)^{-1},\\
T_3^2(\zeta)=1+0.59\zeta+0.65 \zeta^2,
\eea
where $\zeta_i\equiv k/k_i$, with  $k_i$s being the modes  entering the horizon according to Fig.\ref{fig3} and are derived as
\bea
k_{\rm eq}&=&7.1\times 10^{-2}\Omega_m h^2 {\rm Mpc^{-1}},\\
k_{\Phi}&=&1.7\times 10^{14}\left(\frac{g_{*s}(T_\Phi)}{106.75}\right)^{1/6}\left(\frac{T_\Phi}{10^7 \rm GeV}\right){\rm Mpc^{-1}}, \\
k_{\Phi R}&=&1.7\times 10^{14} \kappa^{2/3}\left(\frac{g_{*s}(T_\Phi)}{106.75}\right)^{1/6}\left(\frac{T_\Phi}{10^7 \rm GeV}\right){\rm Mpc^{-1}},
\eea
and 
\bea
k_R=1.7\times 10^{14}\kappa^{-1/3}\left(\frac{g_{*s}(T_R)}{106.75}\right)^{1/6}\left(\frac{T_R}{10^7 \rm GeV}\right){\rm Mpc^{-1}}
\eea
with $T_\Phi\simeq \left(\frac{90}{\pi^2 g_*}\right)^{1/4} \sqrt{\Gamma_N^\Phi \tilde{M}_{Pl}}$. Given the above set of equations and using $\kappa$ from Eq.\ref{entrp},  we now evaluate Eq.\ref{GWeq} for different benchmark values listed in Table \ref{t1}. Throughout the paper we shall use $r=0.06$, $k_*=0.01\rm Mpc^{-1}$ and $h=0.7$. Note also that, by construction, the phase transition responsible for RH neutrino masses happens in the radiation domination after the first reheating, i.e., $T_R>T_c$. With the parameter space of interest, we have $v_\Phi>T_c$, and therefore for simplicity, we shall use only one energy scale $T_R=v_\Phi$ to compute the GW spectrum. In the left panel of Fig.\ref{fig4}, we show the GW spectrum for the following benchmark points:
\bea
{\rm Set~ A}\hspace{3.5cm}\nonumber \\
{\rm ~~~~~~BP1: }~~T_\Phi = 92.9~ {\rm GeV}, ~~\kappa \simeq 2.8\times 10^3,~~n_T=0.9,\nonumber\\
{\rm ~~~~~~BP2: }~~T_\Phi = 9.29~ {\rm GeV}, ~~\kappa \simeq 2.8\times 10^4~~n_T=0.8,\nonumber\\
{\rm ~~~~~~BP3: }~~T_\Phi = 0.929 ~{\rm GeV}, ~~\kappa \simeq 2.8\times 10^5,~~n_T=0.7.\nonumber
\eea
\bea
{\rm Set~ B}\hspace{3.5cm}\nonumber \\
{\rm ~~~~~~BP3: }~~T_\Phi = 16.51 ~{\rm GeV}, ~~\kappa \simeq 4.96\times 10^2,~~n_T=0.6.\nonumber
\eea
 \begin{figure}
\includegraphics[scale=.47]{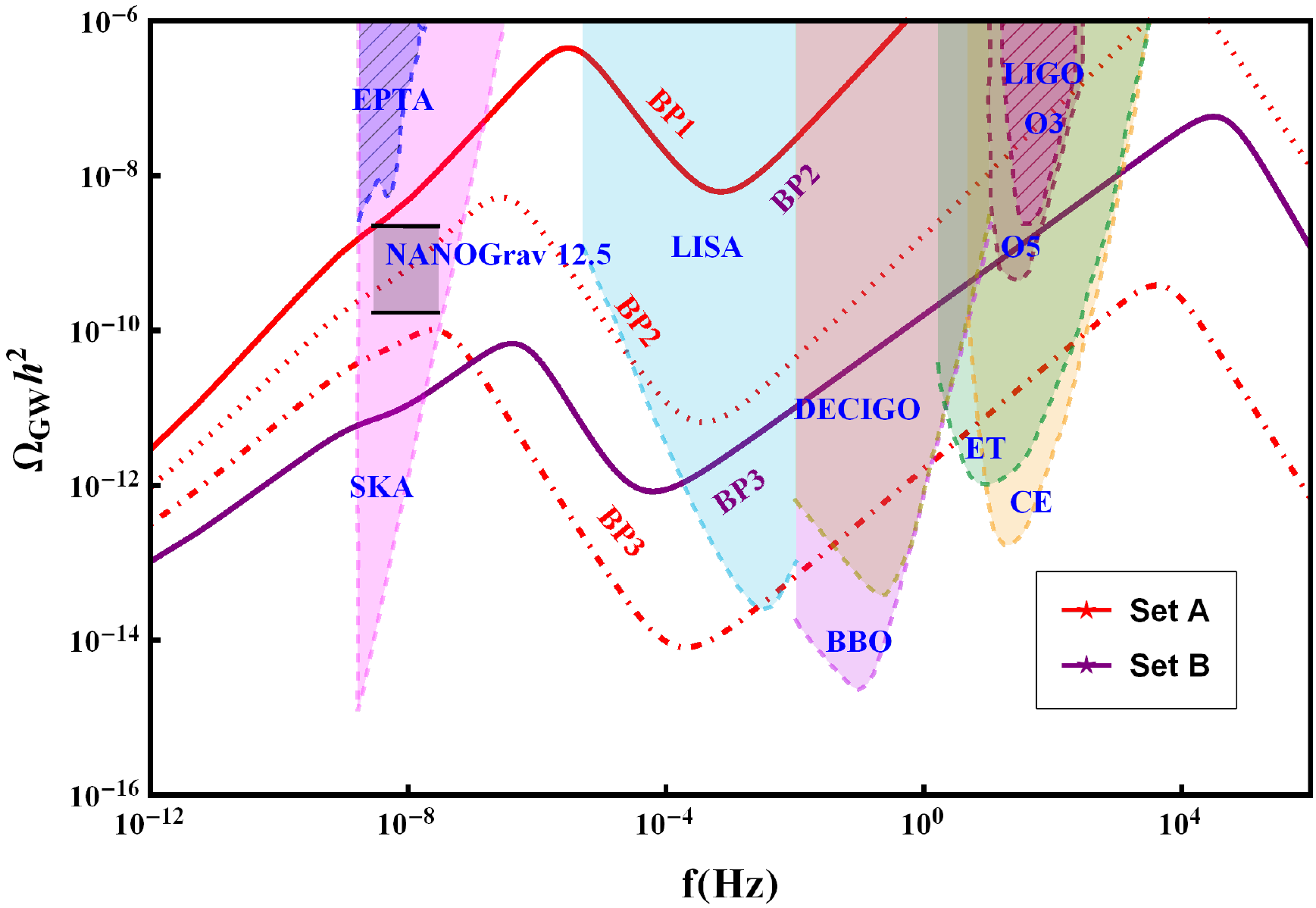}\includegraphics[scale=.47]{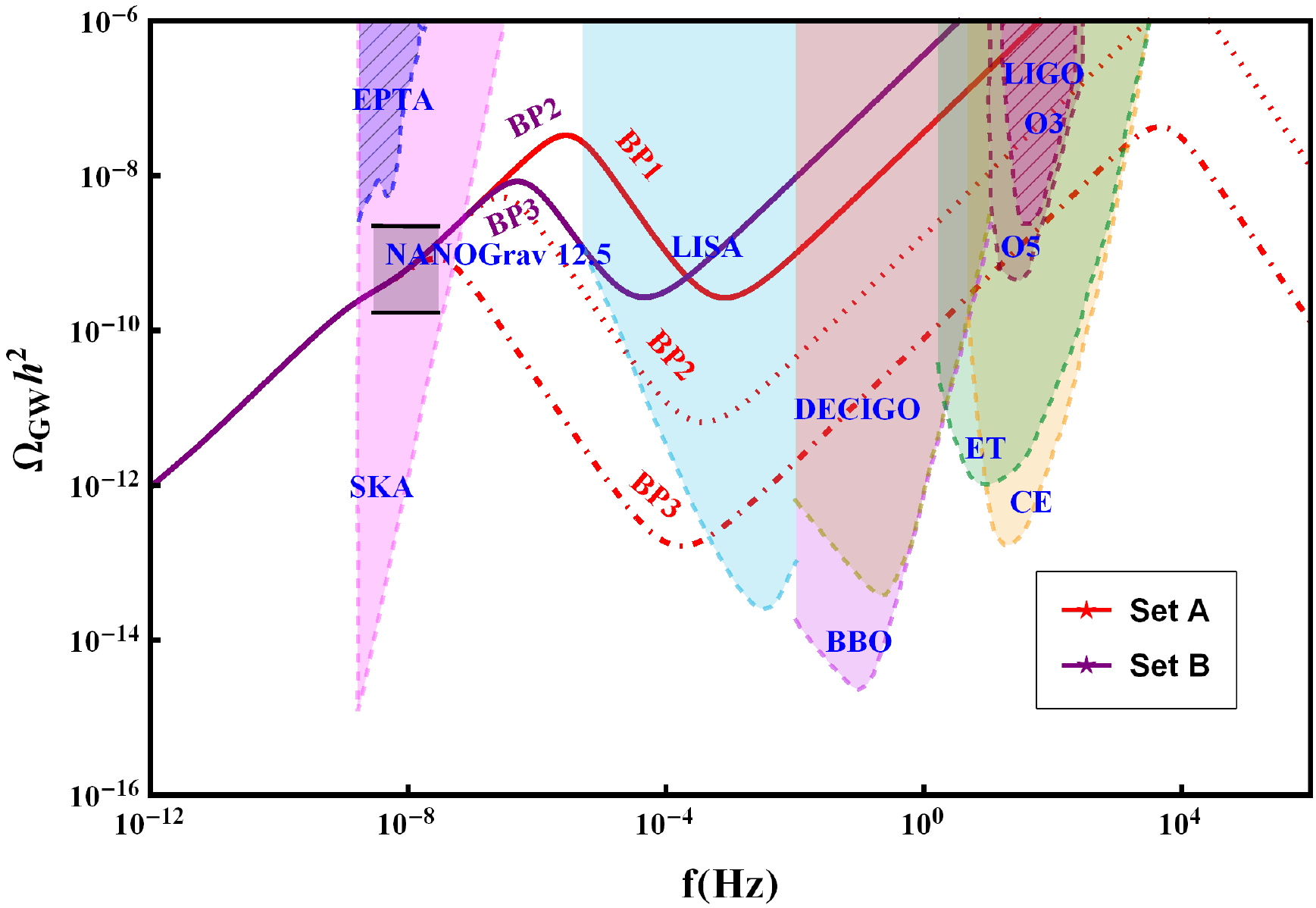}
\caption{Left panel: The GW spectrum for different spectral indices. Set A: BP1 $\Rightarrow n_T=0.9$, BP2 $\Rightarrow n_T=0.8$, BP3 $\Rightarrow n_T=0.7$. Set B: BP3 $\Rightarrow n_T=0.6$. Right Panel: The GW spectrum for $n_T=0.8$. In each panel, there are four curves because BP1 in Set A and BP2 in Set B are the same.}\label{fig4}
\end{figure}
A few comments are in order regarding this figure. First, due to the entropy production by $\Phi$, the spectrum becomes red in the middle, making the overall signal a double-peak spectrum. More significant the entropy production, the lesser the overall amplitude of the signal and the stronger the red-tilt in the middle. Second, although we claim it as the imprint of leptogenesis, we restrict ourselves to the level of RH neutrino masses $M_i$. Nonetheless, depending on the leptogenesis model, a realistic analysis requires the computation of the baryon asymmetry incorporating the effects of entropy production and identifying the relevant parameter space corresponding to the observed baryon asymmetry. For example, one can reproduce figure-5 of Ref.\cite{lep11} including the entropy production and superimposing the GW sensitivity curves on those from the particle physics experiments. Because this article focuses only on the GWs, we shall leave this analysis for a future work. Finally, as the spectrum spans a wide range of frequencies with a distinctive spectral shape, a low-frequency and a complimentary high-frequency measurement of the predicted GWs would be an exciting way to test the signal for a specific value of the spectral index $n_T$. In this regard, we would like to contextualize the recent finding of the NANO-Grav PTA experiment, which reported strong evidence of a stochastic common-spectrum process across 47 millisecond pulsars in the 12.5 years of data \cite{nanograv}. Although the data set does not show any evidence for quadrupolar spatial correlation described by the Hellings-Downs curve \cite{Hellings:1983fr}, the detection, if genuine, opens up a luring possibility to test and constrain particle physics and cosmological models such as the one we discuss here. Intriguingly, evidence of a common-spectrum process has also been reported in the recent data sets of the Parkes Pulsar Timing Array (PPTA) collaboration \cite{ppta} and the European Pulsar Timing Array (EPTA) collaboration \cite{epta}. We find that $n_T\simeq 0.8$ would suffice to be consistent with the NANO-Grav's findings\footnote{For a more accurate and exhaustive fit to the NANO-Grav data with inflationary BGWs, please see Refs.\cite{bn1,bn2,bn3,bn4}. }. In the right panel of Fig.\ref{fig4}, we plot the same benchmark points, but now all of them with $n_T=0.8$. From this figure, we may generally conclude that if we detect inflationary BGWs at low frequencies such as the NANO-Grav one, the amplitude and spectral shapes of the BGWs at higher frequencies can be used to test and constrain LSL models.\\

Notice also that the predicted amplitude could be large at higher frequencies, and  saturate the  BBN bound on effective neutrino species  \cite{Peimbert:2016bdg} as well as the constraint on the SGWB  by the  LIGO \cite{LIGOScientific:2016jlg}. 
\begin{figure}
\includegraphics[scale=.6]{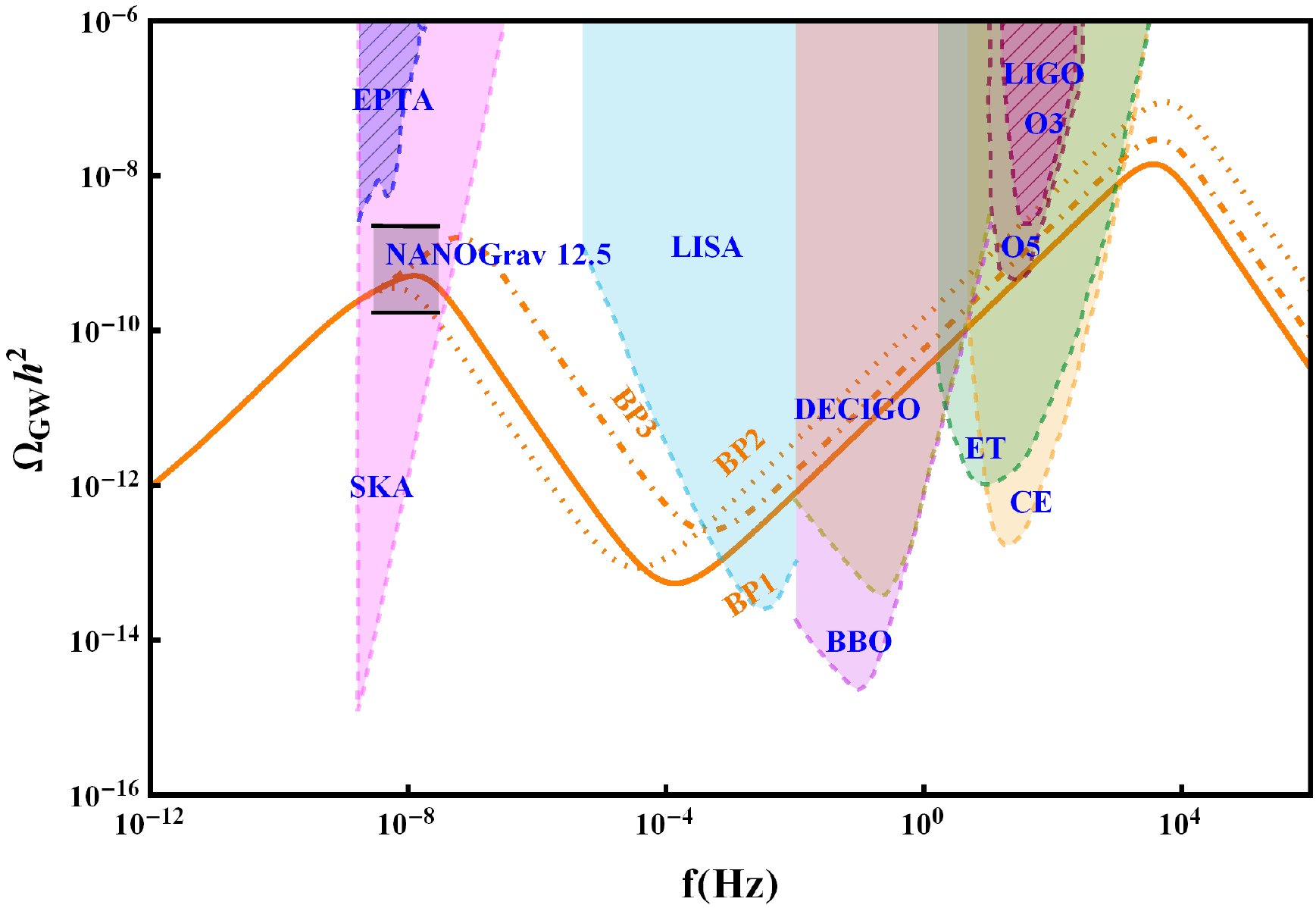}
\caption{The GW spectrum for the benchmark values in the Set C. }\label{figc}
\end{figure}
\begin{table}
	\begin{center}
		\begin{tabular}
			{ |p{2.45cm}||p{1.45cm}|p{2.45cm}|p{2.45cm}|p{1.45cm}|p{2.45cm}|p{2.45cm}| }
			\hline
			\multicolumn{7}{|c|}{Set C: $v_\Phi=10^{13}{\rm GeV}$} \\
			\hline
			Parameters & $g^\prime$  & $\Gamma_N^\Phi/{\rm GeV}$ & $T_c/{\rm GeV}$& $M_i/{\rm GeV}$ & $H (T_c)/{\rm GeV}$&$m_{\Phi}/{\rm GeV}$\\
			\hline
			BP1 & $10^{-3}$  & $3.2\times 10^{-19}$ & $6.32\times 10^{11}$ & $1.5$ & $5.93\times 10^{5}$& $4.47\times 10^{8}$ \\
			BP2 & $10^{-3.5}$  & $1.01\times 10^{-19}$ & $3.55\times 10^{11}$ & $2$ & $1.88\times 10^{5}$& $7.95\times 10^7$\\
			BP3 & $10^{-2.7}$  & $6.42\times 10^{-18}$ & $8.92\times 10^{11}$ & $4$ & $1.18\times 10^{6}$ & $1.26\times10^9$\\
			\hline
		\end{tabular}
	\end{center}
\end{table}
\bea
{\rm Set~ C}\hspace{3.5cm}\nonumber \\
{\rm ~~~~~~BP1: }~~T_\Phi = 0.46~ {\rm GeV}, ~~\kappa \simeq 5.6\times 10^5,~~n_T=0.8,\nonumber\\
{\rm ~~~~~~BP2: }~~T_\Phi = 0.26~ {\rm GeV}, ~~\kappa \simeq 1.7\times 10^5~~n_T=0.8,\nonumber\\
{\rm ~~~~~~BP3: }~~T_\Phi =2.08~{\rm GeV}, ~~\kappa \simeq 3.5\times 10^5,~~n_T=0.8.\nonumber
\eea
 This happens for most of the benchmark points in Fig.\ref{fig4} (see Fig.\ref{figc} and the corresponding benchmarks in Set C, which do not violate the BBN and LIGO bounds). The BBN constraint reads
\bea
\int_{f_{\rm low}}^{f_{\rm high}} f^{-1}df\Omega_{GW}(f)h^2\lesssim 5.6\times 10^{-6}\Delta N_{\rm eff},
\eea
where $\Delta N_{\rm eff}\lesssim 0.2$. The lower limit of the integration is the frequency corresponding to the mode entering the horizon at the BBN epoch, and we take it as $f_{\rm low}\simeq 10^{-10}$ Hz. On the other hand, the upper limit corresponds to the highest frequency of the GWs determined by the Hubble rate at the end of inflation $f_{\rm high}=a_{\rm end}H_{\rm end}/2\pi$. They are different for different benchmark points, but we find that a global choice of $f_{\rm high}\simeq 10^5$ Hz does not alter the result significantly. 
Therefore, we use $f_{\rm high}=10^5$ Hz to derive the BBN constraint for all values of $g^\prime$ and $M_i$. We take into account the LIGO constraint in a much simpler way. We consider a reference frequency $f_{\rm LIGO}=25$ Hz and discard the GWs having amplitude more than $2.2\times 10^{-9}$ \cite{KAGRA:2021kbb}, i.e., we allow the spectrum having $\Omega_{GW}(25{\rm Hz})~h^2\lesssim 2.2\times 10^{-9}$. The constraints are shown in Fig.\ref{fig5} on the $g^\prime-M_i$ plane. The light green and blue regions correspond to those amplitudes that would contradict BBN and LIGO bounds. Although both of them put a severe constraint on the parameter space, ruling out heavier RH neutrino masses, note that they also represent the fact that if the NANO-Grav result is due to the inflationary GWs, future LIGO runs, or more generally, the detectors at the interferometer scales can test and constrain LSL mechanisms for $M_i\lesssim 30 \pm \delta M_i$ GeV (cf. Fig.\ref{fig5}), where $\delta M_i$ should be obtained by taking into account NANO-Grav-$2\sigma$.  The upper bound could be $\mathcal{O}(100)~{\rm GeV}$ for $v_\Phi\sim 10^{14}$ GeV.\\

We conclude this section by pointing out the 
issue that in the future, if GW detectors find such inflationary double-peak GWs, we might not be able to 
discriminate the  LSL mechanisms from any other intermediate matter domination scenarios. Broadly, there is a couple of resolution to this problem. First, let us stress again that the LSL mechanisms are also testable in particle physics experiments \cite{Alekhin:2015byh,Blondel:2014bra,Antusch:2016vyf} which can complement the results obtained from the GW detectors and are therefore distinguishable. Second, if the LSL  mechanisms are constructed in a way that we do here, they can have their own source of GWs from cosmic strings, as discussed recently in the context of seesaw, e.g., in Refs. \cite{Dror:2019syi,Samanta:2020cdk,Datta:2020bht,Samanta:2021zzk,Borah:2022byb,Huang:2022vkf}. We address the latter option in the next section and discuss how  the GWs from cosmic strings \cite{ng1,ng2} can complement the BGWs.

\begin{figure}
\includegraphics[scale=1.2]{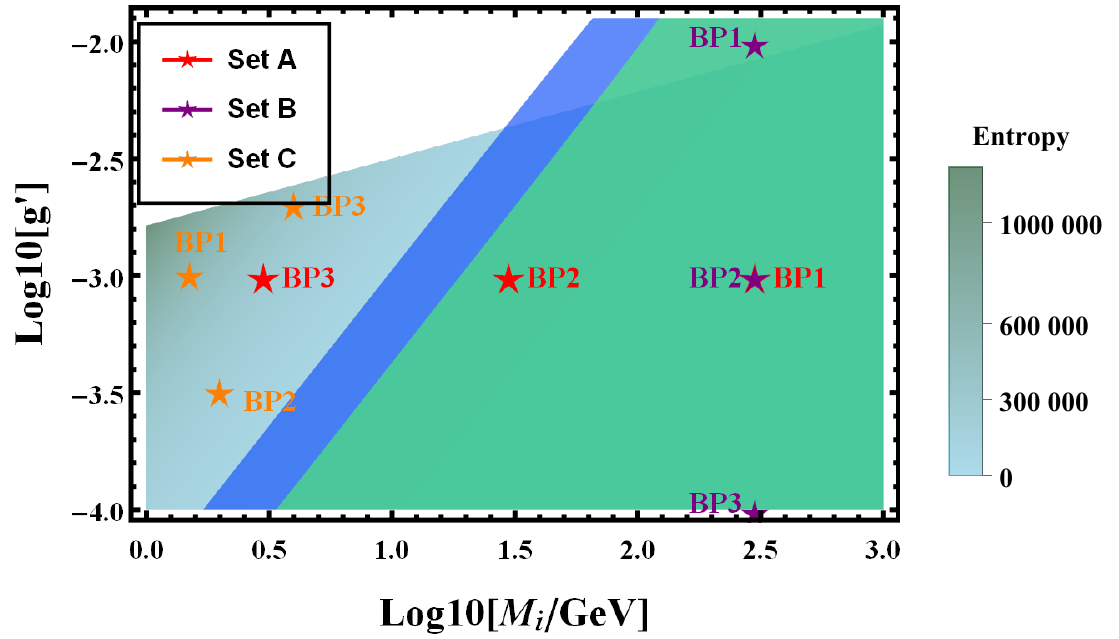}
\caption{Constraints on the $g^\prime-M_i$ parameter space from BBN (light green) and LIGO (blue).  These regions represent the GWs violating BBN and LIGO bounds. }\label{fig5}
\end{figure}

\section{Cosmic strings complementarity}\label{s4}
 Because the vacuum manifold of $U(1)$ is not simply connected \cite{cs1,cs2,cs3}, in this mechanism, cosmic strings appear as topological defects after the spontaneous breaking of the $U(1)$ leading to the massive RH neutrinos. After the formation, the strings get randomly distributed in space forming close loops plus a network of horizon-size long strings~\cite{ls1,ls2}. Two segments of long strings cross, inter-commute, and form loops. Long strings are described by a correlation length $L=\sqrt{\mu/\rho_\infty}$, with $\rho_\infty$ being the long string energy density and $\mu $ being the string tension: defined as $\mu=\pi v_\Phi^2 h\left( \frac{\lambda}{2 {g^\prime}^2}\right)$. The quantity $h$ is a slowly varying function of its arguments with $h\left( \frac{\lambda}{2 {g^\prime}^2}=1\right)\simeq 1$. On the other hand, for $\lambda\ll 2  {g^\prime}^2$, it becomes $h\left( \frac{\lambda}{2 {g^\prime}^2}\right)\simeq \left({\rm ln } ~\frac{2 {g^\prime}^2}{\lambda}\right)^{-1}$ \cite{Hill:1987qx}. Therefore in the present analysis, the string tension is given by $\mu=\pi v_\Phi^2 \left({\rm ln}~\frac{2}{g^\prime}\right)^{-1}$ for $\lambda={g^{\prime}}^3$. \\

Due to the strong interaction with the thermal plasma~\cite{fric}, the motion of a string network gets damped. After the damping phase, the network oscillates to enter a scaling evolution phase, characterized by stretching of the correlation length due to the cosmic expansion and the fragmentation of the long strings into loops. These loops oscillate independently and produce GWs or particle radiation~\cite{cs4,cs5,cs6}. Between these two competing dynamics exists an attractor solution, namely, the scaling regime~\cite{scl1,scl2,scl3}. In this regime, the characteristic length $L$ scales as cosmic time $t$, and therefore,  with a constant  tension, one has $\rho_\infty\propto t^{-2}$. In this way,  the network tracks any cosmological background energy density $\rho_{bg}\propto t^{-2}$ with a small constant proportional to $G\mu$, and does not dominate the energy density of the Universe, unlike any other cosmic defects.\\

There are evidence of scaling solutions in the cosmic string network simulations (see Refs.~\cite{cs7,cs8,scl1,scl2}), and therefore, we assume that the network is in the scaling regime while computing the GW spectrum. The time evolution of a radiating loop of initial size $l_i=\alpha t_i$ is described as $l(t)=\alpha t_i-\Gamma G\mu(t-t_i)$, where $\Gamma\simeq 50$~\cite{cs4,cs6} and $\alpha\simeq 0.1$ ~\cite{cs9,cs10}. The total energy loss from a loop is decomposed into a set of normal-mode oscillations with frequencies $f_j=2j/l_j=a(t_0)/a(t)f$, where $j=1,2,3...j_{max}$ ($j_{max}\rightarrow\infty$). The $j$th mode GW density parameter is given by \cite{cs9} 
\bea
\Omega_{GW}^{(j)}(f)=\frac{2kG\mu^2 \Gamma_j}{f\rho_c}\int_{t_{osc}}^{t_0} \left[\frac{a(t)}{a(t_0)}\right]^5 n\left(t,l_j\right)dt,\label{gwf1}
\eea
where $n\left(t,l_j\right)$ is the scaling loop number density which we calculate from the Velocity-dependent-One-Scale (VOS) model\footnote{It is argued that compared to the numerical simulations \cite{cs9,cs10}, the VOS model overestimates the number density of the loops by an order of magnitude.  It happens because the VOS model considers all the loops are of equal size at production. Nonetheless, there could be a realistic distribution of $\alpha$. Numerical simulations found that only 10$\%$ of the energy of the long string network goes to the large loops ($\alpha\simeq 0.1$) while the rest  goes to the highly boosted smaller loops that are less significant for the GWs. To be consistent with the simulations, we  include a normalisation factor $\mathcal{F}_\alpha\sim 0.1$ in Eq.\ref{gwf1} \cite{vos3}.} \cite{vos1,vos2,vos3}.  For a general equation of state parameter  $\omega$, the number density $n_\omega\left(t,l_k\right)$ is obtained as
\bea
n_\omega(t,l_{j}(t))=\frac{A_\beta}{\alpha}\frac{(\alpha+\Gamma G \mu)^{3(1-\beta)}}{\left[l_k(t)+\Gamma G \mu t\right]^{4-3\beta}t^{3\beta}},\label{genn0}
\eea
where $\beta=2/3(1+\omega)$ and $A_\beta =5.4$ for $w=1/3$, whereas, it is $0.39$ for $\omega=0$ \cite{vos3}. The quantity $\Gamma_j$ is given by $\Gamma_j=\frac{\Gamma j^{-\delta}}{\zeta(\delta)}$, with $\delta=4/3$ for loops containing cusps \cite{cuki}.  The Eq.\ref{gwf1} is valid only for $t_i>t_{osc} = {\rm Max}\,\left[t_F,t_\text{fric}\right]$, where $t_F$ and $t_{\rm fric}$ are network formation time and the time when damping ends. \\

\begin{figure}
	\includegraphics[scale=1.1]{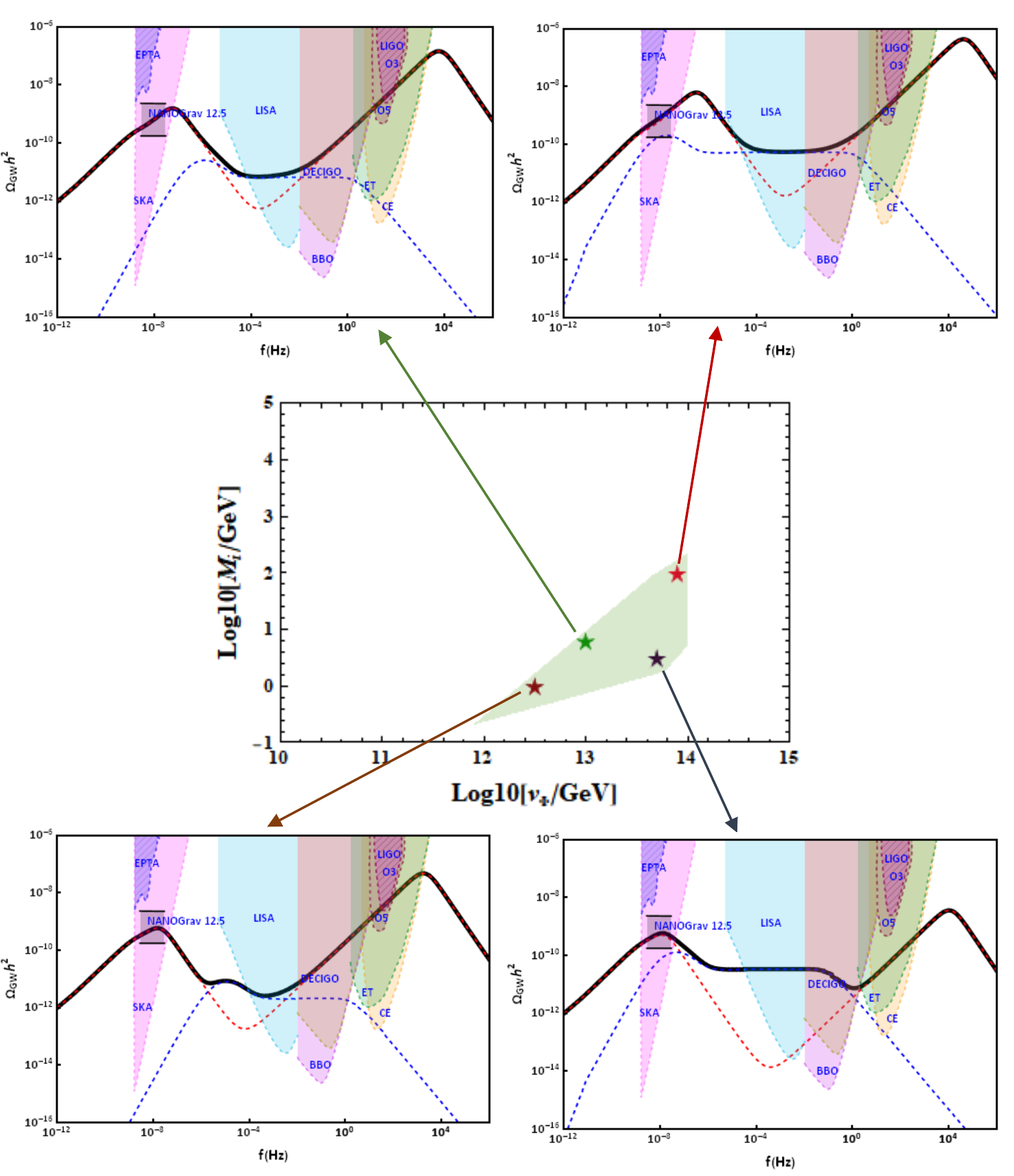}
	\caption{All these figures are produced with $n_T=0.8$ and $g^\prime=10^{-3}$. Black solid lines: Total spectrum. Red dashed lines: Blue-tilted GWs with entropy production in LSL mechanisms. Blue dashed lines: GWs from cosmic strings. }  \label{fig6}
\end{figure}
 \begin{figure}
	\includegraphics[scale=1.1]{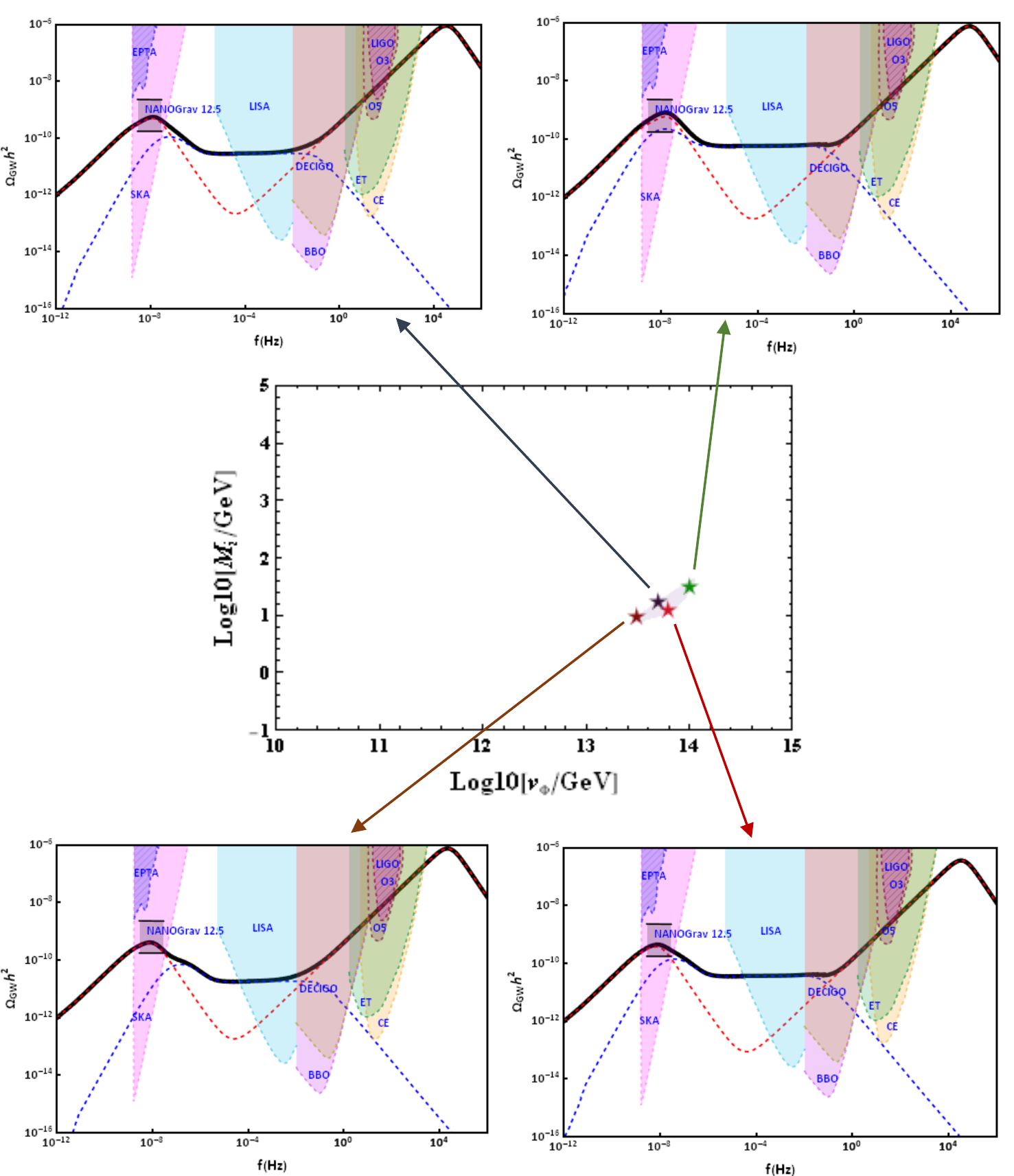}
	\caption{All these figures are produced with $n_T=0.8$ and $g^\prime=10^{-4}$. Black solid lines: Total spectrum. Red dashed lines: Blue-tilted GWs with entropy production in LSL mechanisms. Blue dashed lines: GWs from cosmic strings. } \label{fig7}
\end{figure}
Without any intermediate matter dominated epoch, the GWs arising from Eq.\ref{gwf1} can be described with a peak at a low frequency owing to the GW radiation from the loops  produced in the radiation epoch and decay in the standard matter epoch, plus a plateau at high frequency \cite{cs9,cs10,vos3,Sousa:2020sxs} :
\bea
\Omega_{GW}^{1,~plt}(f)=\frac{128\pi G\mu}{9\zeta(\delta)}\frac{A_r}{\epsilon_r}\Omega_r\left[(1+\epsilon_r)^{3/2}-1\right], \label{flp1}
\eea
that arises due to the loop dynamics only in the radiation epoch. In Eq.\ref{flp1}, 
$\epsilon_r=\alpha/\Gamma G\mu \gg 1$, $A_r\simeq 5.4$  and $\Omega_r\sim 9\times 10^{-5}$. \\

In the presence of a matter-dominated epoch before the most recent radiation epoch, the above description remains the same; barring, the plateau breaks at a high frequency $f_\Delta$, beyond which the spectrum falls as $\Omega_{GW}^{1}(f)\sim f^{-1}$. The frequency $f_\Delta$ can be calculated analytically and is given by \cite{cs11}  
\bea
f_\Delta=\sqrt{\frac{8}{\alpha\Gamma G\mu}}t_\Delta^{-1/2}t_0^{-2/3}t_{\rm eq}^{1/6}\simeq \sqrt{\frac{8 z_{\rm eq}}{\alpha\Gamma G\mu}}\left(\frac{t_{\rm eq}}{t_*}\right)^{1/2}t_0^{-1},\label{br0}
\eea
where $z_{\rm eq}\simeq 3387$ is the red-shift at the standard matter-radiation equality that takes place at time $t_{\rm eq}$, and $t_\Delta<t_{\rm BBN}$ is the time when the matter domination ends. Because we denote the decay temperature of $\Phi$ as $T_\Phi$, we rename $t_\Delta$ and $f_\Delta$ as $t_\Phi^{CS}$ and $f_\Phi^{CS}$ respectively. Furthermore, we shall focus on the results only for the $j=1$ mode, but it can be easily generalized for the infinite number of modes \cite{Datta:2020bht, spb2}.\\

The numerical results of this section are based on the following considerations: I) We stick to $n_T=0.8$ to be consistent with NANO-Grav, since we consider it as a measurement at low frequencies, and taking it at face value, we aim to study the signatures of LSL on the BGWs at higher frequencies. II) To distinguish LSLs from other intermediate matter domination scenarios, we look for spectral distortions at higher frequencies caused by the GWs from cosmic strings. A detectable spectral distortion at higher frequencies would imply $\Omega_{ BGW}\left(f_{\rm dip} \equiv f_{\Phi R}\right)<\Omega_{GW}^{\rm strings}\left(f_{\rm dip} \equiv f_{\Phi R}\right)$. III) A valid overall GW spectrum must not contradict the BBN and LIGO bounds. In Fig.\ref{fig6} (Fig.\ref{fig7}) we show the numerically extracted parameter spaces on the $v_\Phi-M_i$ plane for $g^\prime=10^{-3(-4)}$, which comply with all the above considerations. We also mark four benchmarks to show the expected overall GW spectrum (thick black curves). Note now the following features of Fig.\ref{fig6} and Fig.\ref{fig7}.\\

$\bullet$ For a given $v_\Phi$, there is an absolute upper bound on $M_i$; otherwise, the GWs saturate the BBN and LIGO bounds. Large $M_i$s are disallowed for  smaller values of $v_\Phi$. In this region, the produced entropy is significantly less. Therefore, for a large spectral index $n_T(=0.8)$, the GW amplitudes contradict the BBN and the LIGO bounds easily.\\

$\bullet$ The parameter space never exhibits $\Omega_{ BGW}\left(f_{\rm dip}\right)>\Omega_{GW}^{\rm strings}\left(f_{\rm dip}\right)$. Otherwise, the BGWs always dominate the spectrum, making the existence of cosmic strings irrelevant in the discussion. Unlike Fig.\ref{fig4}, in the presence of GWs from cosmic strings, the total spectrum (thick black curves) flattens at the mid-band, e.g., within the sensitivity range of LISA and DECIGO\footnote{Note that the first peak may fall in the desert region of $\mu$Hz frequencies. Interestingly, there are recent works that explore the possibility of detecting SGWB in the $\mu$Hz region with  Lunar Laser Ranging (LLR) \cite{blas1,blas2}.  }. The flatness becomes more prominent for the smaller values of $M_i$ because for a fixed $v_\Phi$,  a small $M_i$ corresponds to larger entropy production, and therefore compared to the cosmic string spectrum (dashed blue lines), the inflationary spectrum (dashed red lines) becomes significantly sub-dominant at the mid-bands. Note also that in most cases, the GWs from cosmic strings become red at a frequency $f_\Phi^{CS}$ such that no detectable spectral distortion is imprinted on the total spectrum. Therefore only the cosmic string loops which originate and decay in the second radiation epoch, after the $\Phi$ decays, can complement the inflationary GWs by flattening the total spectrum at the mid-bands. \\  

$\bullet$ However, for a large  $v_\Phi$ and a small $M_i$, the high-frequency parts of the signal may show the effect of $\Phi$-domination on the GWs from cosmic strings (Fig.\ref{fig6}, bottom-right). In this region, the entropy production is large enough to suppress BGW amplitudes significantly, making the effect of $\Phi$-domination visible on the cosmic string radiated GWs. Beyond $f_\Phi^{CS}$, the spectrum falls as $f^{-1}$ until the  BGWs take over at the higher frequencies. \\
\begin{figure}
	\includegraphics[scale=.75]{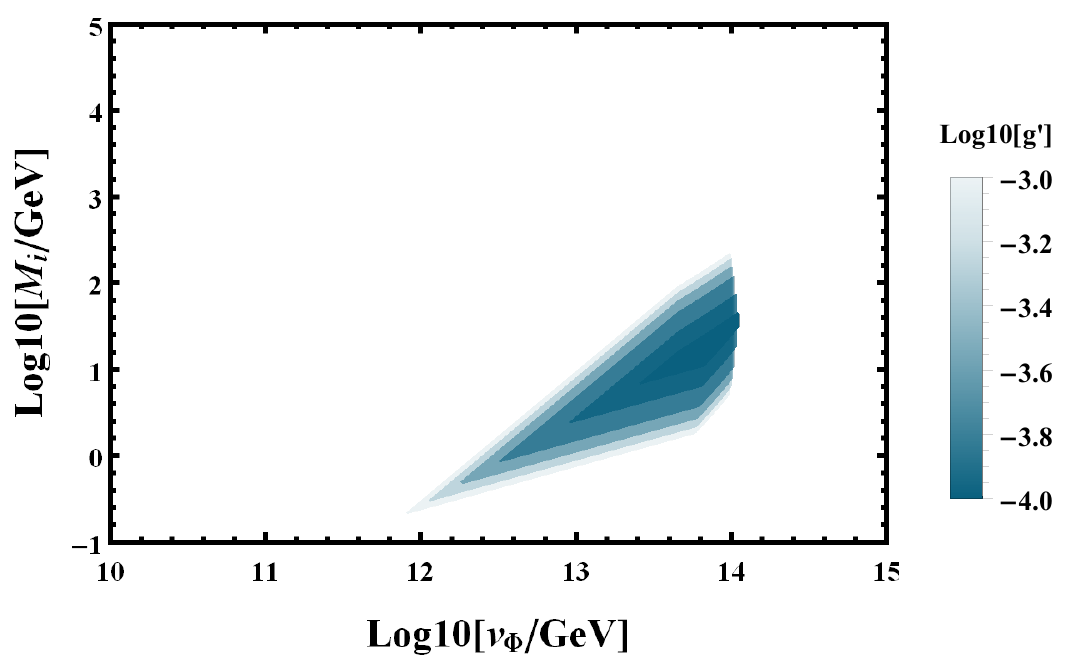} \includegraphics[scale=.75]{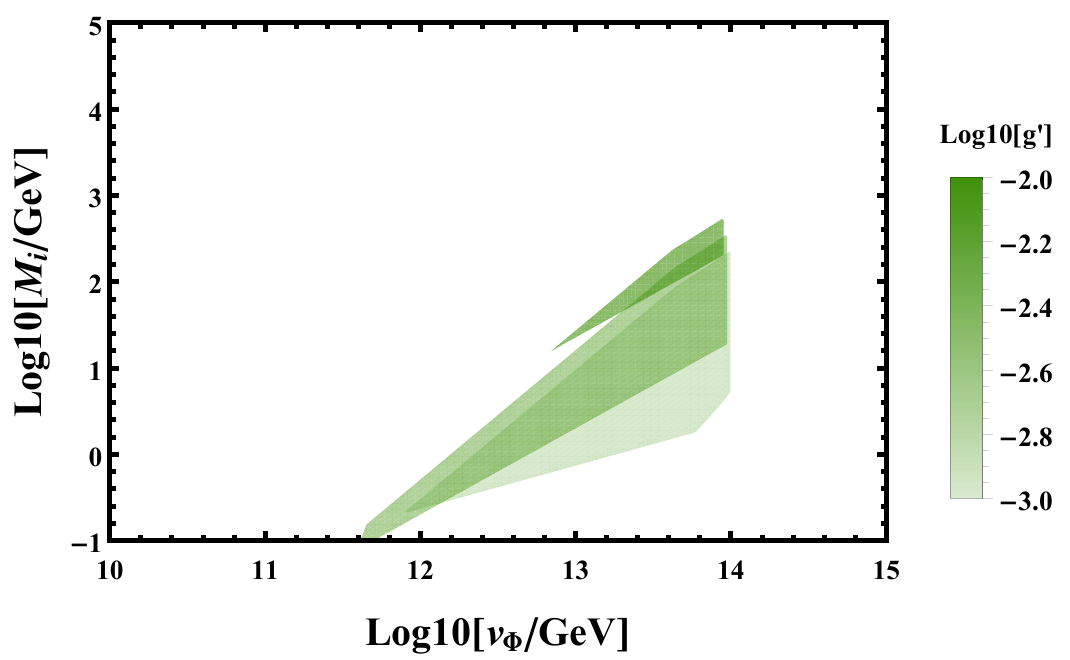}
	\caption{Parameter space allowed for combined GW-tomography of low-scale-leptogenesis for different values of gauge coupling $g^\prime$. Maximally allowed parameter space corresponds to $g^\prime \simeq 10^{-3}$. The parameter space shrinks as $g^\prime$ approaches $10^{-4}$ (left) and $10^{-2}$ (right).}  \label{fig8}
\end{figure}

In Fig.\ref{fig8}, we show the $M_i-v_\Phi$ plot for different values of $g^\prime$ represented by the color gradients.  While a value $g^\prime\simeq 10^{-3}$ corresponds to the largest allowed parameter space,  as we go either side of $g^\prime\simeq 10^{-3}$, it shrinks enormously. The reason is that a larger value; $g^\prime\simeq 10^{-2}$, would strongly violate the condition $\Gamma_{N}^\Phi\gtrsim \Gamma_{ffV}$, whereas, for a smaller value; $g^\prime\simeq 10^{-4}$, the produced entropy is significantly small so that the BBN and the LIGO bounds are strongly violated.   We conclude by pointing out that we work with  unconventional values of $g^\prime$ and $\lambda$, which in general are taken as $g^\prime,~\lambda \simeq 1$ \cite{partrad1,partrad2} in the numerical simulations of cosmic string networks. In the present scenario, for  $v_\Phi\gtrsim 10^{14}$ GeV, the string-width $\delta_w\simeq 1/\sqrt{\lambda}v$ constitutes a considerable fraction of the horizon $H(T_c)^{-1}$ , and   become too thick to be treated as Nambu-Goto strings that predominantly  radiate GWs \cite{ng1,ng2}. Therefore,  the combined GW-tomography of LSL mechanisms works better for $v_\Phi\lesssim 10^{14}$ GeV.

\section{Summary}\label{s5}
We study the possibility of probing low-scale leptogenesis (LSL) mechanisms with inflationary and blue-tilted gravitational waves (GWs). The setup for the LSL mechanism in our study is based on Ref.\cite{Blasi:2020wpy}, wherein a long-lived scalar field generates the required small right-handed (RH) neutrino masses ($M_i$) for LSLs. Due to the large lifetime, the scalar field dominates the Universe's energy density until the most recent radiation domination before the Big-Bang Nucleosynthesis (BBN) takes over. The RH neutrino masses determine the duration of the non-standard cosmic epoch. Such an RH neutrino mass-dependent non-standard cosmic epoch leads to GW-spectral distortions via entropy production, which we claim as the signatures of LSLs. Broadly, there are two significant consequences if the blue-tilted GWs pass through such a cosmic phase. First, depending on the RH neutrino masses, inflationary models with a sizeable tensor-blue-tilt become viable even if the reheating temperature is large. Second, one typically expects a double-peak GW spectrum with the amplitude plus the locations of the peaks depending on the tensor spectral index and LSL model parameters. Unless we fix the spectral index, the study possibly remains academic.
Nonetheless, the recent finding of GW-alike common-spectrum processes by the Pulsar Timing Arrays (PTAs) motivates us to fix the spectral index. In this case, the signatures of LSL on such inflationary GWs become testable. Although the LIGO and BBN severely constrain the model parameter space, we show that if the PTA results are due to blue-tilted inflationary GWs, then future detectors at the interferometer scales have the potential to test and constrain LSL mechanisms with $M_i\lesssim 10^2$ GeV. \\

We finally show that the GWs from cosmic strings that originate naturally due to the phase transition leading to RH neutrino masses can complement the blue-tilted GWs and make the overall GW spectrum distinct. In the presence of cosmic strings, an LSL mechanism in our setup would predict a peak-plateau-peak spectrum instead of a peak-dip-peak spectrum. However, the parameter space of such a spectrum is restrictive and vanishes as the gauge coupling disperses from $g^\prime \sim 10^{-3}$.
\section*{acknowledgement}
R. S is supported  by the  project 
MSCA-IF IV FZU - CZ.02.2.69/0.0/0.0/$20\_079$/0017754 and acknowledges European Structural and Investment Fund and the Czech Ministry of Education, Youth and Sports. We thank Kai Schmitz
for a helpful discussion.
\appendix
\section{Analytical expression for  $g_{*0(s)}(T_{k,\rm in})$}\label{a1}
The analytical expression of $g_{*0(s)}(T_{k,\rm in})$  used in Eq.\ref{fuk} is given by
\bea
g_{*0(s)}(T_{k,\rm in})=g_{*0}\left(\frac{A+{\rm tanh~k_1}}{A+1}\right)\left(\frac{B+{\rm tanh~k_2}}{B+1}\right),
\eea
where 
\bea
A=\frac{-1-10.75/g_{*0(s)}}{-1+10.75/g_{*0(s)}},~~B=\frac{-1-g_{max}/10.75 }{-1+g_{max}/10.75}, 
\eea
and 
\bea
k_1=-2.5~{\rm log}_{10}\left(\frac{k/2\pi}{2.5\times 10^{-12}{\rm Hz}}\right), ~~k_2=-2.0~{\rm log}_{10}\left(\frac{k/2\pi}{6.0\times 10^{-9}{\rm Hz}}\right).
\eea
{}
\end{document}